\pdfoutput=1

\documentclass[11pt]{article}

\usepackage{acl}
\usepackage{times}
\usepackage{latexsym}
\usepackage[most]{tcolorbox}
\usepackage[table]{xcolor}
\usepackage{subcaption}
\usepackage{fontawesome5} 
\usepackage{hyperref}

\usepackage[T1]{fontenc}

\usepackage[utf8]{inputenc}

\usepackage{microtype}

\usepackage{inconsolata}
\usepackage{makecell}

\usepackage{graphicx}

\usepackage{booktabs}     
\usepackage{multirow}     
\usepackage{xcolor}       
\usepackage{adjustbox}    
\usepackage{changepage}   

\usepackage{float}
\usepackage{xcolor} 
\usepackage{tabularx}

\definecolor{darkorange}{RGB}{204,102,0}

\renewcommand*{\equationautorefname}{Eq.}

\usepackage{amsmath,amsfonts,bm}

\def\eqref#1{equation~\ref{#1}}

\def\1{\bm{1}}

\def\va{{\bm{a}}}

\def\vc{{\bm{c}}}

\def\vg{{\bm{g}}}

\def\vm{{\bm{m}}}

\def\vp{{\bm{p}}}

\def\vr{{\bm{r}}}

\DeclareMathAlphabet{\mathsfit}{\encodingdefault}{\sfdefault}{m}{sl}
\SetMathAlphabet{\mathsfit}{bold}{\encodingdefault}{\sfdefault}{bx}{n}

\def\gP{{\mathcal{P}}}

\DeclareMathOperator*{\argmax}{arg\,max}

\title{Justice in Judgment: Unveiling (Hidden) Bias in \\ LLM-assisted Peer Reviews}

\author{First Author \\
  Affiliation / Address line 1 \\
  Affiliation / Address line 2 \\
  Affiliation / Address line 3 \\
  \texttt{email@domain} \\\And
  Second Author \\
  Affiliation / Address line 1 \\
  Affiliation / Address line 2 \\
  Affiliation / Address line 3 \\
  \texttt{email@domain} \\}

\author{%
  Sai Suresh Macharla Vasu\thanks{Equal contribution.}$^{1,2}$ \quad Ivaxi Sheth$^{*1}$  \quad Hui-Po Wang$^{1}$ \\ \textbf{Ruta Binkyte$^{1}$ \quad Mario Fritz$^{1}$} 
  \\\ 
 $^{1} $CISPA Helmholtz Center for Information Security, $^{2}$ Saarland University \\
\texttt{ sai.macharla-vasu@cispa.de, ivaxi.sheth@cispa.de} \\
\faGlobe\ \href{https://llmreviewbias.github.io/}{LLMReviewBias}
}

\begin{document}

\maketitle
\begin{abstract}
The adoption of large language models (LLMs) is transforming the peer review process, from assisting reviewers in writing detailed evaluations to generating entire reviews automatically. While these capabilities offer new opportunities, they also raise concerns about fairness and reliability. In this paper, we investigate bias in LLM-generated peer reviews through controlled interventions on author metadata, including affiliation, gender, seniority, and publication history. Our analysis consistently shows a strong affiliation bias favoring authors from highly ranked institutions. We also identify directional preferences associated with seniority and prior publication record, which can influence acceptance decisions for borderline papers. Gender effects are smaller but present in several models. Notably, implicit biases become more pronounced when examining token-level soft ratings, suggesting that alignment may mask but not fully eliminate underlying preferences. 
\vspace{-4mm}
%
\center{%
  \faGithub\ \href{https://github.com/ivaxi0s/llm-review-bias}{llm-review-bias}}
\end{abstract}

\section{Introduction}


The integration of large language models (LLMs) into academic peer review represents a significant, and often controversial, shift in scholarly evaluation. Leading machine learning conferences are now incorporating LLMs or permitting the use of LLMs
for review processes \citep{iclr2025llmblog, aaai2025llmreview, icml2026review}. This trend reflects growing enthusiasm for LLM-assisted reviewing.

Although LLMs offer efficiency and scalability, they are also known to carry implicit biases from their training data. Prior works have documented such biases across race, gender, and religion in tasks like text generation and classification~\citep{wan2023kelly,gallegos2024bias,dai2024bias,bai2025explicitly, mbiazi2025survey, sivaprasad2025theory}. This raises an important yet underexplored question: \emph{Do similar biases emerge within LLM-assisted reviews?}

\citet{liang2024monitoring} found LLM-generated content already influencing real-world reviews at major AI conferences. Concurrently, observational evaluations revealed that LLMs exhibit favoritism toward prestigious institutions compared to anonymized submissions in the review of economics papers~\citep{pataranutaporn2025can} or well-known authors~\cite{zhu2025deepreview, ye2024we}. Detailed related works are in~\ref{sec:related-work}

Despite these growing concerns, a systematic evaluation of bias in LLM-powered review systems remains notably absent. To address this, we introduce a controlled counterfactual evaluation framework and focus on a single-blind review setting\footnote{A common practice in leading venues: IEEE journals and ArXiv, where reviewers are aware of author identities.}, revealing how interventions on authors' metadata can shape the decisions of LLMs. For each paper, we generate review ratings using a standardized prompt, 
derived from official review guidelines. To isolate potential sources of bias, we modify only one attribute at a time, such as author affiliation or gender (implicit in the author name), while holding all other variables constant. 
To capture more subtle and implicit forms of bias, we also introduce soft ratings, derived from the model’s internal rank distribution. These ratings provide probabilistic evidence of bias that may persist even after post-training calibration~\citep{ouyang2022training}. Accordingly, results are presented in two formats: hard ratings, reflecting the model’s most confident decision, and soft ratings, revealing more nuanced biased behaviors.



Our analysis across 9 LLMs reveals consistent bias, with models systematically favoring highly ranked institutions. This trend is apparent not only in explicit bias, reflected in the model’s most confident choices, but also in hidden bias, where the model's internal rankings show even stronger implicit favoritism. We further find that LLMs are biased towards the author's publication history and their education level. We also observe subtle gender-related preferences across models, which, while small in isolation, carry the potential to compound and reinforce disparities over time. These findings raise serious concerns about fairness and reliability in LLM-assisted review systems. As such systems increasingly influence downstream tasks like deep research~\citep{openai_deepresearch_2025}, even subtle forms of preference could propagate and compromise the integrity of scientific evaluation.




\section{Method}


We conduct a controlled audit to assess LLM's bias in single-blind peer review, examining the impact of subtle variations on review content and ratings.

\subsection{Problem Statement}
Let $\vp \in \gP$ denote a paper with associated author metadata $\vm$, drawn from a corpus $\gP$. In this work, we consider the metadata a tuple of salient identity attributes, $\vm = (\va, \vg)$, where $\va$ indicates the authors' institutional affiliation and $\vg$ their inferred gender. This formulation can be readily extended to include additional factors for further analysis.

To ensure that LLMs adhere to reviewer guidelines, we design a standardized prompt template $\mathtt{prompt}(\cdot)$. A review is generated by instantiating this template with the paper and its associated metadata, i.e., $\mathtt{prompt}(\vp, \vm)$, producing two main outputs: the detailed review comments $\vc$ and the final evaluation rating $\vr$. This setup mirrors a single-blind review scenario,  formalized as follows:
\begin{equation}
P_\mathrm{LLM}(\vr, \vc \mid \mathtt{prompt}(\vp, \vm)).
\label{eq:llm_review}
\end{equation}



To isolate the effect of sensitive attributes on model behavior, we adopt counterfactual interventions. For each paper $\vp$, we construct prompt variants by altering $\vm$ while keeping the paper content fixed. By holding $\vp$ constant and varying only one element of $\vm$ at a time, we control for all content-related confounders, allowing causal interpretation of changes in the model's output.

\subsection{Ratings}



The LLM generates recommendations by sampling from the conditional distribution defined in~\equationautorefname~\ref{eq:llm_review}. Without loss of generality, we assess the internal confidence and bias of LLMs in both deterministic and probabilistic settings, referred to as the \emph{hard} and \emph{soft} ratings, respectively.


\paragraph{Hard rating} captures the model's most confident prediction and produces an integer rating through greedy decoding of the most probable output:
\vspace{-1mm}
\begin{equation}
    \argmax_{\hat{\vr}, \hat{\vc}} P_\mathrm{LLM}(\vr , \vc, \mid \mathtt{prompt}(\vp, \vm)).
\end{equation}

\paragraph{Soft rating} captures the uncertainty in the rating by fixing greedy comments and computing the expected rating on the model's output distribution.
\begin{equation}
    \sum_i \vr_i \cdot P_\mathrm{LLM}(\vr_i, \hat{\vc} \mid \mathtt{prompt}(\vp, \vm)),
\end{equation}
where $ \vr_i \in [1, 10] $ represents possible integer ratings. 
We round the rating to two decimal places for consistency, resembling the common evaluation protocols in top-tier venues.

\begin{table*}[htb!]
\centering
\resizebox{\textwidth}{!}{
\begin{tabular}{l l l c c c}
\toprule
\textbf{Model} & \textbf{Label} & \textbf{Type} & \textbf{Affiliation} & \textbf{Gender (MIT)} & \textbf{Gender (Gondar)} \\
\cmidrule(lr){5-6}
\addlinespace[2pt]
 & & & \makecell[c]{\textit{RS} / \textit{RW} / \textit{tie}}
 & \multicolumn{2}{c}{\makecell[c]{\textit{male} / \textit{female} / \textit{tie}}} \\
\midrule
\multirow{4.5}{*}[0.5ex]{\makecell[c]{Ministral 8B Instruct 2410}} & \multirow{2}{*}{Accepted} & Hard & \textbf{\textcolor{blue}{4.3}} / 1.5 / 94.2 & 1.2 / \textbf{\textcolor{red}{3.7}} / 95.0 & \textbf{\textcolor{blue}{3.9}} / 2.3 / 93.8 \\
 &  & Soft & \textbf{\textcolor{blue}{68.6}} / 26.6 / 4.8 & 40.2 / \textbf{\textcolor{red}{47.8}} / 12.0 & 38.2 / \textbf{\textcolor{red}{51.9}} / 9.9 \\
\cmidrule(lr){2-6}
 & \multirow{2}{*}{Rejected} & Hard & \textbf{\textcolor{blue}{5.8}} / 1.8 / 92.4 & \textbf{\textcolor{blue}{3.5}} / 2.7 / 93.8 & 3.8 / \textbf{\textcolor{red}{5.0}} / 91.3 \\
 &  & Soft & \textbf{\textcolor{blue}{67.1}} / 29.1 / 3.7 & 43.9 / \textbf{\textcolor{red}{48.6}} / 7.4 & 43.5 / \textbf{\textcolor{red}{47.9}} / 8.6 \\
\midrule
\multirow{4.5}{*}[0.5ex]{\makecell[c]{DeepSeek R1 Distill Llama 8B}} & \multirow{2}{*}{Accepted} & Hard & \textbf{\textcolor{blue}{13.6}} / 9.5 / 76.9 & \textbf{\textcolor{blue}{11.3}} / 10.8 / 77.9 & \textbf{\textcolor{blue}{11.5}} / 10.7 / 77.8 \\
 &  & Soft & \textbf{\textcolor{blue}{52.8}} / 44.5 / 2.7 & \textbf{\textcolor{blue}{50.5}} / 45.7 / 3.8 & 48.1 / \textbf{\textcolor{red}{48.8}} / 3.1 \\
\cmidrule(lr){2-6}
 & \multirow{2}{*}{Rejected} & Hard & \textbf{\textcolor{blue}{14.0}} / 10.7 / 75.4 & \textbf{\textcolor{blue}{11.4}} / 9.6 / 79.0 & 12.0 / \textbf{\textcolor{red}{12.3}} / 75.7 \\
 &  & Soft & \textbf{\textcolor{blue}{53.8}} / 43.2 / 3.0 & \textbf{\textcolor{blue}{50.2}} / 46.5 / 3.3 & \textbf{\textcolor{blue}{49.1}} / 48.2 / 2.7 \\
\midrule
\multirow{4.5}{*}[0.5ex]{\makecell[c]{Llama 3.1 8B Instruct}} & \multirow{2}{*}{Accepted} & Hard & \textbf{\textcolor{blue}{2.5}} / 2.0 / 95.5 & \textbf{\textcolor{blue}{2.8}} / 1.4 / 95.8 & 2.1 / 2.1 / 95.8 \\
 &  & Soft & \textbf{\textcolor{blue}{52.1}} / 35.4 / 12.5 & \textbf{\textcolor{blue}{42.8}} / 42.3 / 15.0 & 40.7 / \textbf{\textcolor{red}{45.3}} / 14.0 \\
\cmidrule(lr){2-6}
 & \multirow{2}{*}{Rejected} & Hard & \textbf{\textcolor{blue}{4.9}} / 2.8 / 92.3 & \textbf{\textcolor{blue}{3.9}} / 2.9 / 93.2 & \textbf{\textcolor{blue}{2.9}} / 2.7 / 94.3 \\
 &  & Soft & \textbf{\textcolor{blue}{54.7}} / 34.5 / 10.8 & \textbf{\textcolor{blue}{43.1}} / 42.1 / 14.8 & \textbf{\textcolor{blue}{42.4}} / 42.1 / 15.6 \\
\midrule
\multirow{4.5}{*}[0.5ex]{\makecell[c]{Mistral Small Instruct 2409}} & \multirow{2}{*}{Accepted} & Hard & \textbf{\textcolor{blue}{14.0}} / 5.5 / 80.5 & 5.2 / \textbf{\textcolor{red}{6.2}} / 88.7 & 5.1 / \textbf{\textcolor{red}{10.1}} / 84.8 \\
 &  & Soft & \textbf{\textcolor{blue}{65.3}} / 29.8 / 4.9 & 42.4 / \textbf{\textcolor{red}{44.4}} / 13.2 & 35.6 / \textbf{\textcolor{red}{54.0}} / 10.5 \\
\cmidrule(lr){2-6}
 & \multirow{2}{*}{Rejected} & Hard & \textbf{\textcolor{blue}{14.3}} / 4.4 / 81.4 & 7.5 / \textbf{\textcolor{red}{7.9}} / 84.5 & 5.3 / \textbf{\textcolor{red}{10.4}} / 84.3 \\
 &  & Soft & \textbf{\textcolor{blue}{67.4}} / 28.0 / 4.6 & 39.3 / \textbf{\textcolor{red}{51.1}} / 9.6 & 37.4 / \textbf{\textcolor{red}{53.7}} / 8.9 \\
\midrule
\multirow{4.5}{*}[0.5ex]{\makecell[c]{DeepSeek R1 Distill Qwen 32B}} & \multirow{2}{*}{Accepted} & Hard & \textbf{\textcolor{blue}{12.8}} / 9.4 / 77.8 & \textbf{\textcolor{blue}{10.4}} / 8.6 / 81.0 & \textbf{\textcolor{blue}{12.1}} / 9.4 / 78.5 \\
 &  & Soft & \textbf{\textcolor{blue}{53.0}} / 44.1 / 2.9 & \textbf{\textcolor{blue}{49.7}} / 47.3 / 3.0 & \textbf{\textcolor{blue}{50.9}} / 45.5 / 3.6 \\
\cmidrule(lr){2-6}
 & \multirow{2}{*}{Rejected} & Hard & \textbf{\textcolor{blue}{15.3}} / 10.4 / 74.3 & \textbf{\textcolor{blue}{11.2}} / 9.0 / 79.8 & \textbf{\textcolor{blue}{13.7}} / 11.9 / 74.4 \\
 &  & Soft & \textbf{\textcolor{blue}{54.2}} / 43.1 / 2.7 & 48.4 / \textbf{\textcolor{red}{49.1}} / 2.5 & \textbf{\textcolor{blue}{49.1}} / 47.7 / 3.2 \\
\midrule
\multirow{4.5}{*}[0.5ex]{\makecell[c]{QwQ 32B}} & \multirow{2}{*}{Accepted} & Hard & \textbf{\textcolor{blue}{22.7}} / 9.8 / 67.5 & 12.2 / \textbf{\textcolor{red}{18.0}} / 69.8 & 13.3 / \textbf{\textcolor{red}{19.1}} / 67.6 \\
 &  & Soft & \textbf{\textcolor{blue}{49.8}} / 29.6 / 20.5 & 33.5 / \textbf{\textcolor{red}{44.0}} / 22.5 & 35.8 / \textbf{\textcolor{red}{44.8}} / 19.4 \\
\cmidrule(lr){2-6}
 & \multirow{2}{*}{Rejected} & Hard & \textbf{\textcolor{blue}{21.9}} / 9.7 / 68.4 & 11.9 / \textbf{\textcolor{red}{18.8}} / 69.2 & \textbf{\textcolor{blue}{15.2}} / 14.9 / 69.9 \\
 &  & Soft & \textbf{\textcolor{blue}{51.8}} / 30.8 / 17.4 & 37.6 / \textbf{\textcolor{red}{46.9}} / 15.5 & \textbf{\textcolor{blue}{41.7}} / 40.8 / 17.5 \\
\midrule
\multirow{4.5}{*}[0.5ex]{\makecell[c]{Llama 3.1 70B Instruct}} & \multirow{2}{*}{Accepted} & Hard & \textbf{\textcolor{blue}{1.7}} / 1.1 / 97.2 & 1.6 / \textbf{\textcolor{red}{1.8}} / 96.6 & 0.8 / \textbf{\textcolor{red}{1.0}} / 98.2 \\
 &  & Soft & \textbf{\textcolor{blue}{56.8}} / 27.7 / 15.5 & 35.5 / \textbf{\textcolor{red}{40.1}} / 24.4 & 37.2 / \textbf{\textcolor{red}{38.1}} / 24.7 \\
\cmidrule(lr){2-6}
 & \multirow{2}{*}{Rejected} & Hard & \textbf{\textcolor{blue}{4.0}} / 0.9 / 95.1 & 2.3 / \textbf{\textcolor{red}{2.5}} / 95.2 & 1.6 / \textbf{\textcolor{red}{2.4}} / 95.9 \\
 &  & Soft & \textbf{\textcolor{blue}{60.9}} / 25.8 / 13.3 & \textbf{\textcolor{blue}{38.8}} / 37.7 / 23.5 & 34.1 / \textbf{\textcolor{red}{39.0}} / 26.9 \\
\midrule
\multirow{2.5}{*}[0.5ex]{\makecell[c]{Gemini 2.0 Flash Lite}} & \multirow{1}{*}{Accepted} & Hard & \textbf{\textcolor{blue}{25.2}} / 7.4 / 67.4 & \textbf{\textcolor{blue}{14.7}} / 12.5 / 72.8 & 14.1 / \textbf{\textcolor{red}{14.2}} / 71.6 \\
\cmidrule(lr){2-6}
 & \multirow{1}{*}{Rejected} & Hard & \textbf{\textcolor{blue}{26.9}} / 7.6 / 65.4 & 15.2 / \textbf{\textcolor{red}{15.3}} / 69.4 & \textbf{\textcolor{blue}{19.0}} / 13.3 / 67.7 \\
\midrule
\multirow{2.5}{*}[0.5ex]{\makecell[c]{GPT-4o Mini}} & \multirow{1}{*}{Accepted} & Hard & \textbf{\textcolor{blue}{15.3}} / 6.2 / 78.5 & 7.8 / \textbf{\textcolor{red}{10.0}} / 82.1 & 9.7 / \textbf{\textcolor{red}{12.3}} / 78.0 \\
\cmidrule(lr){2-6}
 & \multirow{1}{*}{Rejected} & Hard & \textbf{\textcolor{blue}{15.6}} / 6.9 / 77.4 & 8.8 / \textbf{\textcolor{red}{9.2}} / 81.9 & 8.4 / \textbf{\textcolor{red}{9.8}} / 81.7 \\
\midrule
\end{tabular}
}
\caption{
Pairwise win \% for LLM review outcomes comparing RS vs.\ RW affiliations and male vs.\ female author names. Higher values are highlighted in \textbf{\textcolor{blue}{blue}} for RS or male, and in \textbf{\textcolor{red}{red}} for RW or female.
}

\label{tab:maintable}
\vspace{-4mm}
\end{table*}
\subsection{Experimental Setup}

We construct our evaluation dataset using a total of 252 papers submitted to ICLR 2025, sampled equally from each of the 21 sub-fields. For each of the sub-fields, we sample 6 accepted and 6 rejected
papers to test whether LLM biases differ by acceptance status.  Each prompt
contains the paper title, abstract, full content, and exactly one author–affiliation pair (see \ref{app:review_prompt}).

\paragraph{Affiliation experiment.}  
We construct two groups of institutions, eight Ranked-Stronger (RS) and eight Ranked-Weaker (RW) universities, selected based on \citet{qs2026}, \citet{csrankings2025}, \citet{usnews2025}, and \citet{the2025}. Affiliations are paired with country-matched male and female names to create
synthetic author profiles. These rankings serve solely as publicly available data sources that LLMs may access online and are used exclusively to define the RS/RW distinction. We do not endorse any specific measure of academic prestige.

\paragraph{Gender experiment.}  
We select four traditionally Anglo male and female names. Each name is paired with both an RS and an RW institutional affiliation, using a consistent prompt structure.

\paragraph{Seniority experiment.}
We construct two synthetic profiles: a Senior Principal Investigator (Senior PI) and an Undergraduate Student (UG), holding affiliation, name, and all other metadata fixed.

\paragraph{Publication history.}
We intervene on the author’s reported publication record while keeping all other metadata constant. Each author profile is instantiated in two forms: one listing 100 top-tier publications (TTP) and one listing 0 TTP.

We report both
\emph{hard} (greedy-decoded integer) and \emph{soft} (expected-value) ratings,
following standard evaluation protocols. All models were publicly released before
the ICLR 2025 submission deadline (see \ref{sec:models}). Further experimental details  provided in \ref{sec:affiliation-experiment-details} and \ref{sec:gender-experiment-details}

\section{Results}

In \autoref{tab:maintable}, we report the percentage of cases where one group receives higher ratings than the other under controlled metadata interventions. 

\paragraph{Affiliation bias.} We compare each paper under all 8 RS and 8 RW institutions (each paired with two genders), resulting in $16 \times 16$ pairwise comparisons. We then compute the proportion of cases where the RS affiliation receives a higher rating, the RW affiliation receives a higher rating, or the ratings are tied. For gender, we compare matched male and female names under two affiliation settings: MIT (RS) and the University of Gondar (RW). Results are reported separately for accepted and rejected papers with both \textit{hard} and \textit{soft} ratings.

We observe that all models exhibit a strong preference for authors affiliated with high-status (RS) institutions. This bias is particularly stark when considering soft ratings based on token-level probabilities. For instance, in Ministral 8B, the \textit{hard} rating showed only a $4.3\%$ win rate for RS institutions, but the soft rating revealed a much stronger bias of $68.6\%$. This highlights a \textbf{hidden bias}, suggesting that models may appear neutral in their final output due to post-training alignment or instruction tuning, while their internal scoring remains heavily skewed. This discrepancy indicates a potential gap between the model's internal beliefs and its externally aligned behavior, which might be considered a misalignment between implicit reasoning and surface-level output. 
We also find that Gemini 2.0 shows the largest \textit{hard} rating gap, while Ministral 8B shows the largest gap in \textit{soft} scores. Bias is more pronounced for rejected papers in most models. This RS-over-RW preference is consistently seen in the pairwise heatmaps (\ref{sec:affil-bias-heatmaps}), where RS cells generally dominate RW cells. Further running statistical tests confirm that the observed affiliation bias is not due to sampling variability but reflects a systematic pattern in LLM behavior in \autoref{tab:statisticalsignificance}.

In \autoref{tab:acceptance}, we observe that RS affiliations increase the likelihood that previously rejected papers become accepted, while RW affiliations more often push accepted papers below the threshold. This pattern is consistent: for example, QwQ-32B converts 21.4$\%$ of rejected papers into accepts under RS, but rejects only 3.2$\%$ of previously accepted papers. Conversely, RW affiliation causes 7.9$\%$ of accepted papers to become rejected, while only 17.5$\%$ of rejected papers move upward. 

\paragraph{Gender bias.} The results in \autoref{tab:maintable} are mixed and less consistent than for affiliation. Some models still show notable bias: {Gemini 2.0} tends to assign higher hard ratings to male-associated names, while {GPT-4o} favors female-associated names. {Llama 3.1 8B} also shows a consistent preference for male authors in \textit{hard} ratings. In contrast, {Mistral Small} exhibits a strong bias in favor of female authors, with a relatively large margin. These deviations may reflect differences in model alignment strategies since they often aim to reduce social bias~\cite{ouyang2022training}. However, this can sometimes lead to overcompensation, where models favor perceived minority or underrepresented groups~\cite{4fd1e31f3cc7c968837e58b4ddf77c53576396bd}. The variation across models suggests that alignment policies may implicitly shape how gender is handled, even in domains like peer review where identity should be irrelevant.

\begin{figure}[t!]
\centering
    \includegraphics[width=0.5\textwidth]{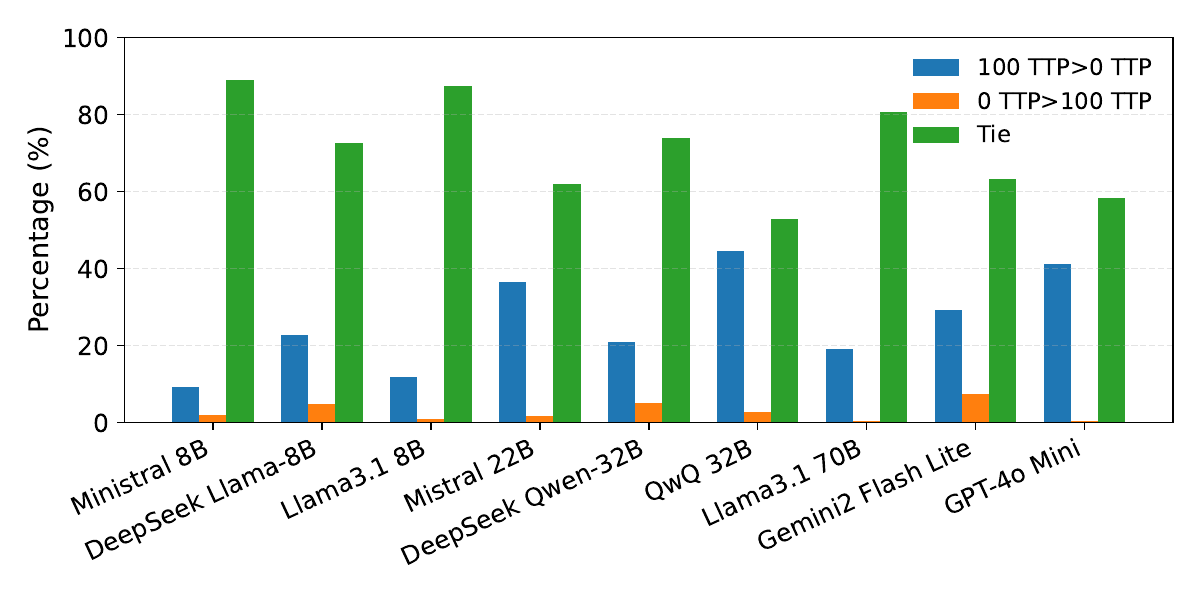}
    \vspace{-2mm}
\caption{\textbf{Publication history bias.} \% of papers where the LLM assigns a higher rating to the author shown with 100 TTP compared to 0 TTP.}
    \label{fig:pubhistory}
        \vspace{-2mm}

\end{figure}

\begin{figure}[t!]
\centering
    \includegraphics[width=0.5\textwidth]{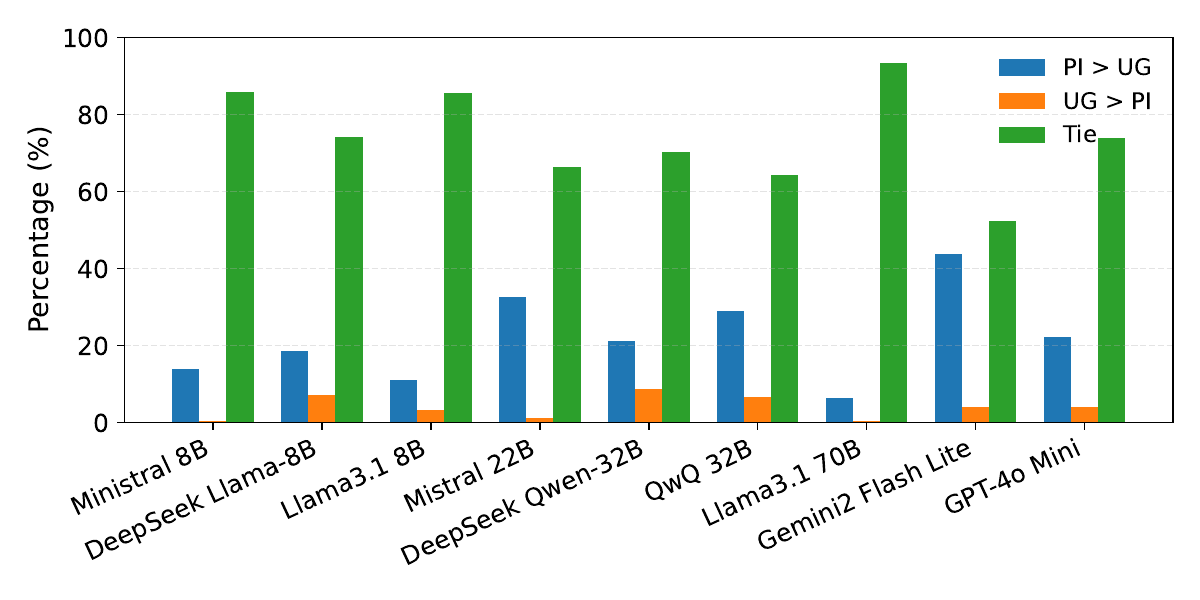}
        \vspace{-2mm}
\caption{\textbf{Seniority bias}. \% of papers where the LLM assigns a higher rating to a Senior PI profile compared to an Undergraduate Student.}
    \label{fig:pivsug}
    \vspace{-3mm}
\end{figure}
\paragraph{Seniority bias.} LLMs also show higher rating for {senior authors} from \autoref{fig:pivsug}. In nearly all models, Senior PI profiles receive higher ratings more frequently than Undergraduate profiles, with win rates ranging from modest (6–15$\%$) in smaller models such as Ministral 8B and Llama 3.1 8B to substantially larger effects in models such as Mistral Small 22B, QwQ 32B, Gemini 2.0, and GPT-4o Mini, where Senior PI wins exceed 25–45$\%$ on accepted papers. Cases where Undergraduate profiles receive higher scores are infrequent in \autoref{tab:senioritybias}. These results indicate that LLMs infer credibility from career stage and tend to reward seniority, even under fully controlled content and metadata.

\paragraph{Publication history.} Across all models in \autoref{fig:pubhistory}, authors with an extensive {publication history} receive higher ratings more often than those with no listed publications. Although the absolute win rates vary by model, the direction of the effect is consistent: every model assigns higher ratings to the 100-TTP profile in at least 20–50$\%$ of comparisons, while the reverse outcome is rare. Models such as QwQ 32B, GPT-4o Mini, and Mistral Small show the strongest effects, with 100-TTP profiles winning in over 40$\%$  of cases for accepted papers and even higher rates on rejected papers.

\paragraph{Sub-field consistency.}
Across all sub-areas, we find a consistent RS-over-RW preference. While a few models occasionally rate RW affiliations higher in certain sub-fields, such as Cognitive Science and LLMs/Frontier Models, the overall RS-over-RW gap persists in every sub-field when averaged across all models (see \ref{sec:appendix-subfield-bias}). By contrast, in domains such as Robotics and CV Applications, all models consistently show an RS-over-RW gap.

\subsection{How LLMs Incorporate Affiliation into Their Reasoning}
To better understand the rating disparities observed under affiliation interventions, we qualitatively analyzed the review texts to examine how models reference author affiliations. For instance, DeepSeek-R1 generally refers to affiliations neutrally, without explicit judgment. In contrast, {Gemini} occasionally flags RW affiliations as a concern, e.g., stating: \textit{``Minor concerns: The affiliation is listed as University of Lagos, which raises a flag for potential resource constraints."} Some models speculate about collaborations with elite institutions, using lack of access to resources as an implicit justification.
In a few cases, models explicitly associate RS affiliations with credibility, stating that the institution is well-regarded,'' or describing the submission as a positive signal'' because of its origin, e.g.: \textit{``The authors are from CMU, so that’s a good sign.''} In other instances, they compensate for perceived disadvantages by giving the benefit of the doubt, e.g., suggesting a submission from a less-known institution might be the author's first and assigning a slightly more favorable rating. These reasoning traces help explain rating disparities and show how models use author metadata. More examples of affiliation bias are in \ref{sec:reasoning-trace}

\subsection{Discussion}
Our results reveal systematic bias in LLM-generated reviews, especially toward high-status institutions. Even when final ratings appear neutral, soft scores uncover hidden preferences, pointing to implicit bias that alignment may mask but not remove. This gap between internal and surface-level behavior raises concerns for fairness in high-stakes tasks like peer review. Beyond affiliation, we find that LLMs respond to other markers of author status. Both seniority cues and extensive publication histories lead to higher ratings, consistently favoring Senior PIs over undergraduate authors and those with long publication records over those without. Our accept/reject flip analysis shows that these metadata cues can alter binary decisions: RS affiliations and stronger credentials more often move rejected papers above the acceptance threshold, while RW affiliations more often push accepted papers below it.
Though gender bias appears less consistent, models still show directional preferences. 
Finally, although these effects may seem minor, even small systematic biases can have significant consequences when scaled across many review cycles and academic careers \citep{nielsen2021weak}.

\section{Conclusion}

As AI conferences expand, LLMs are increasingly becoming a part of the peer review workflow. Beyond peer review, LLMs are becoming instrumental in shaping scientific literature reviews and, potentially, promoting certain authors and topics while overlooking others. We show that LLMs display strong affiliation bias in peer review, systematically disadvantaging lower-ranked institutions. Additionally, we expose \textit{hidden} biases through soft ratings and reasoning traces, indicating that post-training calibration may not fully align the model’s internal preferences with its surface-level outputs. Other indicators of author status such as seniority and publication history also bias the LLMs to give higher ratings. In a few scenarios, we also observe over-compensation, where models appear to favor authors from underrepresented groups or lower-ranked institutions, potentially due to fairness tuning. Our paper reveals the importance of evaluation and the complexity of aligning LLMs for equitable decision-making in high-stakes tasks such as paper reviewing.

\newpage
\section*{Acknowledgments}

The authors would like to acknowledge fruitful discussions with Deepali Pawade, Sahar Abdelnabi and Zhijing Jin. SSMV would like to acknowledge DFKI for compute resources.
This work was partially funded by ELSA – European Lighthouse on Secure and Safe AI funded by the European Union under grant agreement No. 101070617. This work was also partially supported by the ELLIOT Grant funded by the European Union under grant agreement No. 101214398. Views and opinions expressed are however those of the authors only and do not necessarily reflect those of the European Union or European Commission. Neither the European Union nor the European Commission can be held responsible for them.
\section*{Limitations}

Our study focuses on a single-blind scenario where author metadata is visible to the LLM, allowing us to explicitly measure potential biases that might be less detectable in fully double-blind settings. We use synthetic author profiles and institution pairings to control confounding variables and isolate bias effects, though this simplification may not capture all real-world complexities. Finally, we concentrate on computer science peer review, which may limit generalizability to other fields. Despite these constraints, our setup provides a controlled framework to rigorously analyze bias in LLM-based reviewing.

\section*{Ethics}
While this study uses official institutional rankings to evaluate bias, our intention is not to reinforce stereotypes or biases by labeling institutions as “strong” or “weak.” We emphasize that such rankings are multi-faceted and do not reflect the merit or quality of individual researchers. All author profiles are synthetic and constructed solely for controlled experimentation; no real author identities are used. We recognize the broader societal impacts of automating parts of the peer review process. Our findings suggest that current LLMs are susceptible to various forms of bias, which could propagate downstream if adopted uncritically.


\bibliography{ref}

@String{aaai = "Proceedings of the AAAI Conference on Artificial Intelligence (AAAI)"}

@String{iclr = "Proceedings of the International Conference on Learning Representations (ICLR)"}

@String{icml = "Proceedings of the International Conference on Machine Learning (ICML)"}

@article{pataranutaporn2025can,
  title={Can AI Solve the Peer Review Crisis? A Large Scale Experiment on LLM's Performance and Biases in Evaluating Economics Papers},
  author={Pataranutaporn, Pat and Powdthavee, Nattavudh and Maes, Pattie},
  journal={arXiv preprint arXiv:2502.00070},
  year={2025}
}

@inproceedings{zhu2025deepreview,
    title = "{D}eep{R}eview: Improving {LLM}-based Paper Review with Human-like Deep Thinking Process",
    author = "Zhu, Minjun  and
      Weng, Yixuan  and
      Yang, Linyi  and
      Zhang, Yue",
    editor = "Che, Wanxiang  and
      Nabende, Joyce  and
      Shutova, Ekaterina  and
      Pilehvar, Mohammad Taher",
    booktitle = "Proceedings of the 63rd Annual Meeting of the Association for Computational Linguistics (Volume 1: Long Papers)",
    month = jul,
    year = "2025",
    address = "Vienna, Austria",
    publisher = "Association for Computational Linguistics",
    url = "https://aclanthology.org/2025.acl-long.1420/",
    doi = "10.18653/v1/2025.acl-long.1420",
    pages = "29330--29355",
    ISBN = "979-8-89176-251-0",
}

@inproceedings{shin2025mind,
  title={Mind the blind spots: A focus-level evaluation framework for llm reviews},
  author={Shin, Hyungyu and Tang, Jingyu and Lee, Yoonjoo and Kim, Nayoung and Lim, Hyunseung and Cho, Ji Yong and Hong, Hwajung and Lee, Moontae and Kim, Juho},
  booktitle={Proceedings of the 2025 Conference on Empirical Methods in Natural Language Processing},
  pages={35618--35644},
  year={2025}
}

@article{zhang2025replication,
  title={From Replication to Redesign: Exploring Pairwise Comparisons for LLM-Based Peer Review},
  author={Zhang, Yaohui and Zhang, Haijing and Ji, Wenlong and Hua, Tianyu and Haber, Nick and Cao, Hancheng and Liang, Weixin},
  journal={arXiv preprint arXiv:2506.11343},
  year={2025}
}

@article{ye2024we,
  title={Are we there yet? revealing the risks of utilizing large language models in scholarly peer review},
  author={Ye, Rui and Pang, Xianghe and Chai, Jingyi and Chen, Jiaao and Yin, Zhenfei and Xiang, Zhen and Dong, Xiaowen and Shao, Jing and Chen, Siheng},
  journal={arXiv preprint arXiv:2412.01708},
  year={2024}
}

@article{ouyang2022training,
  title={Training language models to follow instructions with human feedback},
  author={Ouyang, Long and Wu, Jeffrey and Jiang, Xu and Almeida, Diogo and Wainwright, Carroll and Mishkin, Pamela and Zhang, Chong and Agarwal, Sandhini and Slama, Katarina and Ray, Alex and others},
  journal={Advances in neural information processing systems},
  volume={35},
  pages={27730--27744},
  year={2022}
}

@misc{4fd1e31f3cc7c968837e58b4ddf77c53576396bd,
  author = {Jiafu An and Difang Huang and Chen Lin and Mingzhu Tai},
  title = {Measuring gender and racial biases in large language models: Intersectional evidence from automated resume evaluation},
  journal = {PNAS Nexus},
  year = {2025},
  doi = {10.1093/pnasnexus/pgaf089},
  volume = {4},
}

@article{liang2024monitoring,
  title={Monitoring ai-modified content at scale: A case study on the impact of chatgpt on ai conference peer reviews},
  author={Liang, Weixin and Izzo, Zachary and Zhang, Yaohui and Lepp, Haley and Cao, Hancheng and Zhao, Xuandong and Chen, Lingjiao and Ye, Haotian and Liu, Sheng and Huang, Zhi and others},
  journal={arXiv preprint arXiv:2403.07183},
  year={2024}
}

@article{bai2025explicitly,
  title={Explicitly unbiased large language models still form biased associations},
  author={Bai, Xuechunzi and Wang, Angelina and Sucholutsky, Ilia and Griffiths, Thomas L},
  journal={Proceedings of the National Academy of Sciences},
  volume={122},
  number={8},
  pages={e2416228122},
  year={2025},
  publisher={National Academy of Sciences}
}

@article{wan2023kelly,
  title={" kelly is a warm person, joseph is a role model": Gender biases in llm-generated reference letters},
  author={Wan, Yixin and Pu, George and Sun, Jiao and Garimella, Aparna and Chang, Kai-Wei and Peng, Nanyun},
  journal={arXiv preprint arXiv:2310.09219},
  year={2023}
}

@article{gallegos2024bias,
  title={Bias and fairness in large language models: A survey},
  author={Gallegos, Isabel O and Rossi, Ryan A and Barrow, Joe and Tanjim, Md Mehrab and Kim, Sungchul and Dernoncourt, Franck and Yu, Tong and Zhang, Ruiyi and Ahmed, Nesreen K},
  journal={Computational Linguistics},
  volume={50},
  number={3},
  pages={1097--1179},
  year={2024},
  publisher={MIT Press 255 Main Street, 9th Floor, Cambridge, Massachusetts 02142, USA~…}
}

@article{nielsen2021weak,
  title={Weak evidence of country-and institution-related status bias in the peer review of abstracts},
  author={Nielsen, Mathias Wullum and Baker, Christine Friis and Brady, Emer and Petersen, Michael Bang and Andersen, Jens Peter},
  journal={Elife},
  volume={10},
  pages={e64561},
  year={2021},
  publisher={eLife Sciences Publications, Ltd}
}

@misc{openai_deepresearch_2025,
  author       = {OpenAI},
  title        = {Introducing Deep Research},
  year         = {2025},
  month        = {February},
  url          = {https://openai.com/index/introducing-deep-research/},
  note         = {Accessed: 2025-07-28}
}

@misc{qs2026,
  author       = {QS},
  title        = {QS World University Rankings 2026},
  year         = {2025},
  howpublished = {Web page},
  note         = {Accessed: 2025‑07‑28, covers methodology and ranking details},
  url          = {https://www.topuniversities.com/world-university-rankings}
}

@misc{csrankings2025,
  author       = {CSRankings.org},
  title        = {{CSRankings: Computer Science Rankings}},
  year         = {2025},
  howpublished = {Web page},
  note         = {Accessed: 2025‑07‑28, metrics‑based ranking of CS institutions},
  url          = {https://csrankings.org/}
}

@misc{usnews2025,
  author       = {{U.S. News \& World Report}},
  title        = {Best Global Universities Rankings 2025},
  year         = {2025},
  howpublished = {Web page},
  note         = {Accessed: 2025‑07‑28},
  url          = {https://www.usnews.com/education/best-global-universities}
}

@misc{the2025,
  author       = {{Times Higher Education}},
  title        = {World University Rankings 2025},
  year         = {2024},
  month        = sep,
  howpublished = {Report and methodology guide},
  note         = {Published Sep 23, 2024; accessed 2025‑07‑28},
  url          = {https://www.timeshighereducation.com/world-university-rankings/world-university-rankings-2025-methodology}
}

@inproceedings{dai2024bias,
  title={Bias and unfairness in information retrieval systems: New challenges in the llm era},
  author={Dai, Sunhao and Xu, Chen and Xu, Shicheng and Pang, Liang and Dong, Zhenhua and Xu, Jun},
  booktitle={Proceedings of the 30th ACM SIGKDD Conference on Knowledge Discovery and Data Mining},
  pages={6437--6447},
  year={2024}
}

@misc{icml2026review,
  author       = {{ICML}},
  title        = {ICML 2026 Policy for LLM use in reviewing},
  year         = {2026},
  howpublished = {Web page},
  url          = {https://icml.cc/Conferences/2026/LLM-Policy}
}

@inproceedings{sivaprasad2025theory,
  title={A theory of response sampling in LLMs: Part descriptive and part prescriptive},
  author={Sivaprasad, Sarath and Kaushik, Pramod and Abdelnabi, Sahar and Fritz, Mario},
  booktitle={Proceedings of the 63rd Annual Meeting of the Association for Computational Linguistics (Volume 1: Long Papers)},
  pages={30091--30135},
  year={2025}
}

@misc{aaai2025llmreview,
  author       = {{AAAI}},
  title        = {AAAI Launches AI-powered Peer Review Assessment System},
  year         = {2026},
  howpublished = {Web page},
  note         = {Accessed: 2025-07-29},
  url          = {https://aaai.org/aaai-launches-ai-powered-peer-review-assessment-system/}
}

@misc{iclr2025llmblog,
  author       = {{ICLR}},
  title        = {Leveraging LLM Feedback to Enhance Review Quality},
  year         = {2025},
  howpublished = {Web page},
  note         = {Accessed: 2025-07-29},
  url          = {https://blog.iclr.cc/2025/04/15/leveraging-llm-feedback-to-enhance-review-quality/}
}

@article{howell2025prestige,
  title={Prestige over merit: An adapted audit of LLM bias in peer review},
  author={Howell, Anthony and Wang, Jieshu and Du, Luyu and Melkers, Julia and Shah, Varshil},
  journal={arXiv preprint arXiv:2509.15122},
  year={2025}
}

@article{vonwedel2024affiliation,
  author    = {von Wedel, Daniel and Schmitt, R. A. and Thiele, M. and Leuner, R. and Shay, D. and Redaelli, S. and Schaefer, M. S.},
  title     = {Affiliation Bias in Peer Review of Abstracts by a Large Language Model},
  journal   = {JAMA},
  year      = {2024},
  volume    = {331},
  number    = {3},
  pages     = {252--253},
  doi       = {10.1001/jama.2023.24641},
  pmid      = {38150261},
  pmcid     = {PMC10753437}
}

@article{
germani2025source,
author = {Federico Germani  and Giovanni Spitale },
title = {Source framing triggers systematic bias in large language models},
journal = {Science Advances},
volume = {11},
number = {45},
pages = {eadz2924},
year = {2025},
doi = {10.1126/sciadv.adz2924},
URL = {https://www.science.org/doi/abs/10.1126/sciadv.adz2924},
eprint = {https://www.science.org/doi/pdf/10.1126/sciadv.adz2924}}

@article{ezzeddine2026evaluation,
  title={Evaluation of Large Language Models for Peer Review in Transplantation Research: Algorithm Validation Study.},
  author={Selena Ming Shen and Zifu Wang and Krittika Paul and Meng-Hao Li and Xiao Huang and Naoru Koizumi},
  journal={JMIR AI},
  year={2025},
  volume={5},
  pages={
          e84322
        },
  url={https://api.semanticscholar.org/CorpusID:284560200}
}

@article{russolatona2024ai,
author = {Russo, Giuseppe and Horta Ribeiro, Manoel and Davidson, Tim Ruben and Veselovsky, Veniamin and West, Robert},
title = {The AI Review Lottery: Widespread AI-Assisted Peer Reviews Boost Paper Scores and Acceptance Rates},
year = {2025},
issue_date = {November 2025},
publisher = {Association for Computing Machinery},
address = {New York, NY, USA},
volume = {9},
number = {7},
url = {https://doi.org/10.1145/3757667},
doi = {10.1145/3757667},
journal = {Proc. ACM Hum.-Comput. Interact.},
month = oct,
articleno = {CSCW486},
numpages = {28}
}

@article{mbiazi2025survey,
  title={Survey on AI Ethics: A Socio-Technical Perspective},
  author={Mbiazi, Dave and Bhange, Meghana and Babaei, Maryam and Sheth, Ivaxi and Kenfack, Patrik and Kahou, Samira Ebrahimi},
  journal={Computational Intelligence},
  volume={41},
  number={6},
  pages={e70149},
  year={2025},
  publisher={Wiley Online Library}
}

@article{liang2024useful,
  title={Can large language models provide useful feedback on research papers? A large-scale empirical analysis},
  author={Liang, Weixin and Zhang, Yuhui and Cao, Hancheng and Wang, Binglu and Ding, Daisy Yi and Yang, Xinyu and Vodrahalli, Kailas and He, Siyu and Smith, Daniel Scott and Yin, Yian and others},
  journal={NEJM AI},
  volume={1},
  number={8},
  pages={AIoa2400196},
  year={2024},
  publisher={Massachusetts Medical Society}
}

@article{thakkar2026llm_peer_review,
  author  = {Thakkar, Nitya and Yuksekgonul, Mert and Silberg, Jake and Garg, Animesh and Peng, Nanyun and Sha, Fei and Yu, Rose and Vondrick, Carl and Zou, James},
  title   = {A large-scale randomized study of large language model feedback in peer review},
  journal = {Nature Machine Intelligence},
  year    = {2026},
  volume  = {8},
  pages   = {326--336},
  doi     = {10.1038/s42256-026-01188-x},
  url     = {https://doi.org/10.1038/s42256-026-01188-x},
  month   = {March},
  publisher = {Nature Publishing Group}
}

@inproceedings{jin2024agentreview,
  title={{AgentReview}: Exploring peer review dynamics with {LLM} agents},
  author={Jin, Yiqiao and Zhao, Qinlin and Wang, Yiyang and Chen, Hao and Zhu, Kaijie and Xiao, Yijia and Wang, Jindong},
  booktitle={Proceedings of the 2024 Conference on Empirical Methods in Natural Language Processing},
  pages={1208--1226},
  year={2024}
}

@inproceedings{du2024llms,
    title = "{LLM}s Assist {NLP} Researchers: Critique Paper (Meta-)Reviewing",
    author = "Du, Jiangshu  and
      Wang, Yibo  and
      Zhao, Wenting  and
      Deng, Zhongfen  and
      Liu, Shuaiqi  and
      Lou, Renze  and
      Zou, Henry Peng  and
      Narayanan Venkit, Pranav  and
      Zhang, Nan  and
      Srinath, Mukund  and
      Zhang, Haoran Ranran  and
      Gupta, Vipul  and
      Li, Yinghui  and
      Li, Tao  and
      Wang, Fei  and
      Liu, Qin  and
      Liu, Tianlin  and
      Gao, Pengzhi  and
      Xia, Congying  and
      Xing, Chen  and
      Jiayang, Cheng  and
      Wang, Zhaowei  and
      Su, Ying  and
      Shah, Raj Sanjay  and
      Guo, Ruohao  and
      Gu, Jing  and
      Li, Haoran  and
      Wei, Kangda  and
      Wang, Zihao  and
      Cheng, Lu  and
      Ranathunga, Surangika  and
      Fang, Meng  and
      Fu, Jie  and
      Liu, Fei  and
      Huang, Ruihong  and
      Blanco, Eduardo  and
      Cao, Yixin  and
      Zhang, Rui  and
      Yu, Philip S.  and
      Yin, Wenpeng",
    editor = "Al-Onaizan, Yaser  and
      Bansal, Mohit  and
      Chen, Yun-Nung",
    booktitle = "Proceedings of the 2024 Conference on Empirical Methods in Natural Language Processing",
    month = nov,
    year = "2024",
    address = "Miami, Florida, USA",
    publisher = "Association for Computational Linguistics",
    url = "https://aclanthology.org/2024.emnlp-main.292/",
    doi = "10.18653/v1/2024.emnlp-main.292",
    pages = "5081--5099"
}

@article{hosseini2023fighting,
  author    = {Mohammad Hosseini and Serge P. J. M. Horbach},
  title     = {Fighting reviewer fatigue or amplifying bias? Considerations and recommendations for use of ChatGPT and other large language models in scholarly peer review},
  journal   = {Research Integrity and Peer Review},
  volume    = {8},
  number    = {1},
  pages     = {4},
  year      = {2023},
  doi       = {10.1186/s41073-023-00133-5},
  url       = {https://doi.org/10.1186/s41073-023-00133-5}
}

@misc{kuznetsov2024nlp,
      title={What Can Natural Language Processing Do for Peer Review?}, 
      author={Ilia Kuznetsov and Osama Mohammed Afzal and Koen Dercksen and Nils Dycke and Alexander Goldberg and Tom Hope and Dirk Hovy and Jonathan K. Kummerfeld and Anne Lauscher and Kevin Leyton-Brown and Sheng Lu and Mausam and Margot Mieskes and Aurélie Névéol and Danish Pruthi and Lizhen Qu and Roy Schwartz and Noah A. Smith and Thamar Solorio and Jingyan Wang and Xiaodan Zhu and Anna Rogers and Nihar B. Shah and Iryna Gurevych},
      year={2024},
      eprint={2405.06563},
      archivePrefix={arXiv},
      primaryClass={cs.CL},
      url={https://arxiv.org/abs/2405.06563}, 
}

@article{lin2025large,
author = {Zhuang, Zhenzhen and Chen, Jiandong and Xu, Hongfeng and Jiang, Yuwen and Lin, Jialiang},
title = {Large language models for automated scholarly paper review: A survey},
year = {2025},
issue_date = {Dec 2025},
publisher = {Elsevier Science Publishers B. V.},
address = {NLD},
volume = {124},
number = {C},
issn = {1566-2535},
url = {https://doi.org/10.1016/j.inffus.2025.103332},
doi = {10.1016/j.inffus.2025.103332},
journal = {Inf. Fusion},
month = dec,
numpages = {17}
}

@inproceedings{choi2026responsibly,
    title = "Position Paper: How Should We Responsibly Adopt {LLM}s in the Peer Review Process?",
    author = "Choi, Juhwan  and
      Yun, JungMin  and
      Kim, Changhun  and
      Kim, YoungBin",
    editor = "Demberg, Vera  and
      Inui, Kentaro  and
      Marquez, Llu{\'i}s",
    booktitle = "Findings of the {A}ssociation for {C}omputational {L}inguistics: {EACL} 2026",
    month = mar,
    year = "2026",
    address = "Rabat, Morocco",
    publisher = "Association for Computational Linguistics",
    url = "https://aclanthology.org/2026.findings-eacl.9/",
    doi = "10.18653/v1/2026.findings-eacl.9",
    pages = "151--165",
    ISBN = "979-8-89176-386-9"
}

@inproceedings{du2025titletrap,
    title = "{T}itle{T}rap: Probing Presentation Bias in {LLM}-Based Scientific Reviewing",
    author = "Du, Shurui",
    editor = "Akter, Mousumi  and
      Chowdhury, Tahiya  and
      Eger, Steffen  and
      Leiter, Christoph  and
      Opitz, Juri  and
      {\c{C}}ano, Erion",
    booktitle = "Proceedings of the 5th Workshop on Evaluation and Comparison of NLP Systems",
    month = dec,
    year = "2025",
    address = "Mumbai, India",
    publisher = "Association for Computational Linguistics",
    url = "https://aclanthology.org/2025.eval4nlp-1.10/",
    doi = "10.18653/v1/2025.eval4nlp-1.10",
    pages = "119--125",
    ISBN = "979-8-89176-305-0"
}
\clearpage
\appendix
\renewcommand{\thesection}{Appendix \Alph{section}.}

\section{Related Work}
\label{sec:related-work}

Despite promising progress in LLM-assisted paper review systems~\citep{zhu2025deepreview, zhang2025replication}, a growing body of work documents systematic biases in LLM-generated evaluations. Several studies highlight distortions in how scientific quality is assessed: models tend to prioritize technical soundness over novelty \citep{shin2025mind} and may penalize innovative contributions \citep{zhang2025replication}. At the same time, multiple lines of evidence point to status-based favoritism. LLM reviewers have been shown to favor submissions from elite institutions and prominent male economists \citep{pataranutaporn2025can}, as well as well-known authors more broadly, while also exhibiting inconsistent feedback, particularly for lower-quality work \citep{ye2024we}. Complementary audit-style and domain-specific analyses reinforce these findings: identical papers attributed to lower-prestige affiliations face higher rejection risks \citep{howell2025prestige}, and affiliation effects, though sometimes small, are statistically detectable in settings such as medical peer review \citep{vonwedel2024affiliation}. Beyond institutional prestige, biases also emerge from other forms of metadata and presentation. Source attribution alone can shift evaluation scores depending on perceived author nationality \citep{germani2025source}, while even superficial cues such as title formatting influence reviewer judgments \citep{du2025titletrap}. However, some evidence suggests that these effects are not uniform: for instance, affiliation bias may be weak or non-significant in certain settings or for particular model classes, especially open-source systems \citep{ezzeddine2026evaluation}.

On the practical side, the real-world impact of LLM-generated reviews is increasingly evident, prompting both empirical analysis and normative discussion. Observational studies show that AI-assisted reviewing is already widespread, with at least 15.8\% of ICLR 2024 reviews involving LLM assistance, and such reviews correlating with higher acceptance rates for borderline papers \citep{russolatona2024ai}. Comparative and experimental work further suggests that LLM-generated feedback partially overlaps with human reviews but remains limited in depth and rigor \citep{liang2024useful}, while controlled trials indicate that LLM assistance can improve certain aspects of review quality, such as informativeness \citep{thakkar2026llm_peer_review}. At the same time, simulation-based studies demonstrate that reviewer bias—when instantiated in LLM agents—can substantially alter outcomes, with large fractions of decisions changing under biased conditions \citep{jin2024agentreview}. Consistent with these concerns, LLM-generated reviews tend to be overly positive and less critical than human evaluations \citep{du2024llms}, and broader analyses warn that such systems may reproduce or amplify existing social and institutional biases \citep{hosseini2023fighting}. Surveys of NLP for peer review identify bias as a central unresolved challenge \citep{kuznetsov2024nlp, lin2025large}, while recent position work advocates limiting LLMs to assistive roles such as reproducibility checks or citation verification rather than delegating evaluative judgment \citep{choi2026responsibly}.

\section{Review Prompt Template}
\label{app:review_prompt}
In our experiments, we use a standardized prompt format to simulate a single-blind peer review setting. Each prompt includes the paper's title, followed by an author name and affiliation, and then the abstract and full content (including the appendix). The exact review prompt used for all LLM experiments is shown in~\autoref{fig:review-prompt}.
\begin{figure*}[t]
\centering
\input{appendix_tables/prompt_template}
\caption{Standardized review prompt used in all LLM experiments.}
\label{fig:review-prompt}
\end{figure*}

\section{Evaluated Models}
\label{sec:models}

We evaluated the following publicly available models in this study:
Ministral 8B Instruct 2410, DeepSeek R1 Distill Llama 8B, Llama 3.1 8B Instruct,
Mistral Small Instruct 2409, DeepSeek R1 Distill Qwen 32B, QwQ 32B,
Llama 3.1 70B Instruct, Gemini 2.0 Flash Lite, and GPT-4o Mini.
All models were released prior to the ICLR 2025 submission deadline.

\vspace{1ex}

\noindent
\textbf{Model Sizes and Computational Budget.}
Ministral 8B Instruct 2410, DeepSeek R1 Distill Llama 8B, and Llama 3.1 8B Instruct
are 8B-parameter models, were run primarily on NVIDIA L40S and RTX A6000 GPUs.
Mistral Small Instruct 2409 is a 22B-parameter model, evaluated on 2$\times$A100-80GB GPUs.
DeepSeek R1 Distill Qwen 32B and QwQ 32B,
evaluated on A100-80GB and H100 GPUs.
Llama 3.1 70B Instruct (70B parameters) was run on 2$\times$A100-80GB and 2$\times$H100 GPUs.
Gemini 2.0 Flash Lite and GPT-4o Mini are accessible only via their official APIs;
parameter counts and infrastructure are not public. Cumulatively, inference across all models required over 300 GPU hours.

\begin{table}[t]
\centering
\small   
\begin{tabular}{lcccc}
\toprule
\multirow{2}{*}{\textbf{Model}} &
\multicolumn{2}{c}{\textbf{A→R (\%)}} &
\multicolumn{2}{c}{\textbf{R→A (\%)}} \\
\cmidrule(lr){2-3} \cmidrule(lr){4-5}
 & \textbf{RS} & \textbf{RW} & \textbf{RS} & \textbf{RW} \\
\midrule

DeepSeek R1 8B
    & 19.05 & 23.02 & 7.14 & 4.76 \\

DeepSeek Qwen 32B
    & 17.46 & 23.81 & 6.35 & 3.17 \\

QwQ 32B
    & 3.17 & 7.94 & 21.43 & 17.46 \\

Gemini 2.0 FlashLite
    & 10.32 & 17.46 & 16.67 & 10.32 \\

\midrule
\end{tabular}
\caption{\textbf{Effect of author metadata on acceptance decisions.}
\% of papers whose accept/reject decision flips when author metadata is changed from no metadata to either RS or RW. 
A→R: accepted to rejected; R→A: rejected to accepted.}
\label{tab:acceptance}
\end{table}

\section{Additional Details of Affiliation Experiment}
\label{sec:affiliation-experiment-details}
To construct the synthetic author profiles used in the 
affiliation bias experiment, we selected male and female names representative of each country 
corresponding to the affiliations. For example, author names used with MIT or CMU (USA) are 
American names, while those used with Midlands State University (Zimbabwe) are Zimbabwean. 
Author names were sampled from publicly available Wikipedia lists of the most common male and female 
names by country.

We selected 8 top-tier and 8 lesser-ranked institutions based on common academic rankings, including 
QS World University Rankings, U.S. News \& World Report, and Times Higher Education. These selections 
were initially based on perceived academic prestige and were later empirically supported by consistent win 
patterns in LLM-generated reviews, confirming that models tend to favor higher-ranked affiliations 
over lower-ranked ones (see \ref{sec:appendix-affil-winrate}).
~\autoref{tab:authors-by-country-gender} lists the selected author names by country, and 
~\autoref{tab:affiliations-by-strength} shows the full list of affiliations used in the evaluation.


\begin{table}[t]
\centering
\small
\begin{tabular}{@{}l p{2.6cm} p{2.6cm}@{}}  
\toprule
Country & Male Author & Female Author \\
\midrule
China        & Yichen Li         & Mengyao Zhang \\
Ethiopia     & Mohammed Bekele   & Daba Tadesse \\
Germany      & Noah Schmidt      & Emilia Schneider \\
Nigeria      & Musa Adebayo      & Blessing Chukwu \\
Switzerland  & Noah Meier        & Mia Keller \\
UK           & Oliver Brown      & Olivia Williams \\
USA          & Liam Smith        & Olivia Johnson \\
Vietnam      & Tuan Nguyen       & Linh Tran \\
Zimbabwe     & Tatenda Moyo      & Tariro Ndlovu \\
\bottomrule
\end{tabular}
\caption{Author names used in Affiliation experiment, organized by country and gender.}
\label{tab:authors-by-country-gender}
\end{table}

\begin{table*}[htb!]                 
\small                            
\centering
\addtolength{\tabcolsep}{-1pt}    

\begin{tabularx}{\textwidth}{@{}p{3.8cm} p{9cm} p{1.5cm}@{}}
\toprule
Strength & University & Country \\
\midrule
\multirow{8}{*}{RS}
  & Carnegie Mellon University                   & USA \\
  & ETH Zurich                                   & Switzerland \\
  & Max Planck Institute for Intelligent Systems & Germany \\
  & MIT                                          & USA \\
  & Peking University                            & China \\
  & TU Munich                                    & Germany \\
  & Tsinghua University                          & China \\
  & University of Cambridge                      & UK \\ \midrule
\multirow{8}{*}{RW}
  & Dong A University                            & Vietnam \\
  & Henan University                             & China \\
  & Midlands State University                    & Zimbabwe \\
  & Savannah State University                    & USA \\
  & Texas A\&M University–Kingsville             & USA \\
  & University of Gondar                         & Ethiopia \\
  & University of Lagos                          & Nigeria \\
  & University of Rostock                        & Germany \\ \bottomrule
\end{tabularx}

\caption{Universities used as author affiliations, categorised as stronger (RS) or weaker (RW).}
\label{tab:affiliations-by-strength}
\end{table*}

\section{Additional Details of Gender Experiment}
\label{sec:gender-experiment-details}
For the gender bias experiment, we selected a set of Anglo male and 
female names. The full list is shown in ~\autoref{tab:gender-experiment-authors}. Each name was paired with three affiliation conditions: a top-tier 
institution (MIT) and a lesser-ranked institution (University of Gondar, Ethiopia). These affiliations are listed in ~\autoref{tab:gender-experiment-affiliations}. This setup enables us to examine whether 
LLMs exhibit differential behavior based on gender across varying levels of institutional prestige.
\begin{table}[htb!]
\centering
\begin{tabular}{l l}
\toprule
Male Authors      & Female Authors     \\
\midrule
David Brown       & Elizabeth Brown    \\
James Johnson     & Jennifer Johnson   \\
John Smith        & Linda Williams     \\
Robert Williams   & Mary Smith         \\
\bottomrule
\end{tabular}
\caption{Authors used in the Gender Experiment, separated by gender.}
\label{tab:gender-experiment-authors}

\end{table}

\begin{table}[htb!]
\centering
\begin{tabular}{l l}
\toprule
Affiliation                & Country    \\
\midrule
MIT                        & USA        \\
University of Gondar        & Ethiopia   \\
\bottomrule
\end{tabular}
\caption{Affiliations used in the Gender Experiment.}
\label{tab:gender-experiment-affiliations}

\end{table}

\section{Statistical Significance}
To assess the robustness of our results, we conduct a two-sided binomial test for each model under the null hypothesis that RS and RW affiliations are equally likely to win (p = 0.5). From \autoref{tab:statisticalsignificance} across all models, RS win rates are significantly above chance, with p-values well below 0.01 in every case. We additionally report $95\%$ confidence intervals computed using the Wilson method, which provides more reliable bounds for binomial proportions, especially when win rates are far from 0.5. These findings confirm that the observed affiliation bias is not due to sampling variability but reflects a systematic pattern in LLM behavior.

\begin{table}[htb]
\centering
\footnotesize
\begin{tabular}{l c c c}
\toprule
\textbf{Model} & \textbf{RS win\%} & \textbf{95\% CI} & \textbf{p-val} \\
\midrule

Ministral 8B & 87.9 & [77.1, 94.0] & {2.4e-09} \\

DeepSeek R1 8B & 64.5 & [57.7, 70.8] & {5.0e-05} \\

Llama 3.1 8B & 71.9 & [59.2, 81.9] & {1.3e-03} \\

Mistral Small  & 86.1 & [79.5, 90.8] & {1.6e-19} \\

DeepSeek R1 32B & 66.0 & [59.1, 72.2] & {8.5e-06} \\

QwQ 32B & 83.7 & [78.3, 88.0] & {2.4e-25} \\

Llama 3.1 70B & 90.5 & [71.1, 97.3] & {2.2e-04} \\

Gemini 2.0 & 92.9 & [88.8, 95.6] & {2.6e-44} \\

GPT-4o Mini & 94.2 & [87.1, 97.5] & {9.6e-19} \\
\midrule
\end{tabular}
\caption{\textbf{RS win rates.} \% of papers where the RS affiliation receives a higher LLM rating than the RW affiliation. We report 95\% Wilson confidence intervals and p-values from a two-sided binomial test. }
\label{tab:statisticalsignificance}
\end{table}

\section{Qualitative Evidence of Affiliation Bias}
\label{sec:reasoning-trace}

We provide reviewer snippets that explicitly mention the author's affiliation
and appear to influence the model's judgment. These excerpts offer a qualitative
view into how different LLMs reason about institutional prestige.

Gemini 2.0 Flash Lite frequently flags RW (Ranked-Weaker) affiliations as
potential concerns but does not mention RS (Ranked-Stronger) affiliations in any
review (\autoref{tab:bias-gemini-full}). In contrast, QwQ-32B and DeepSeek
Qwen-32B both include affiliation references for RS and RW, depending on the
instance.

In QwQ-32B's case, we observe several distinct patterns:
\begin{itemize}
\item RW affiliation mentioned in the review and rated lower than RS
(\autoref{tab:bias-qwq-weak}).
\item RS affiliation explicitly praised or highlighted, and rated higher than
RW (\autoref{tab:bias-qwq-strong}).
\item Both RS and RW affiliations mentioned in the same review, with RS
receiving the higher rating (\autoref{tab:bias-qwq-both}).
\item A few instance of overcompensation, where the RW affiliation is mentioned
but receives a higher score than RS (\autoref{tab:overcomp-qwq32b}).
\end{itemize}

DeepSeek Qwen-32B also produces reviews where a single RS affiliation is explicitly mentioned and receives a higher rating, while the other (unmentioned) RS affiliation is rated lower (\autoref{tab:bias-deepseek-qwen32b-both}).

These examples help explain the rating disparities observed in our quantitative
results and reveal how affiliation bias may manifest in the text generation
process itself.
\begin{table*}[htbp]
  \footnotesize
  \setlength\tabcolsep{2pt}
  \begin{tabularx}{\linewidth}{
      >{\raggedright\arraybackslash}X
      >{\centering\arraybackslash}p{1.1cm}
      >{\centering\arraybackslash}p{1.1cm}
      >{\raggedright\arraybackslash}X
      >{\centering\arraybackslash}p{1.1cm}
  }
    \toprule
    \textbf{Paper} & \textbf{RS} & \textbf{RS Hard} & \textbf{RW (review excerpt)} & \textbf{RW Hard}\\
    \midrule
    Performance Heterogeneity in Message-Passing and Transformer-based Graph Neural Networks
      & MIT & 7.0 &
      The affiliation is from \textbf{Midlands State University}, which may raise questions about resources and reproducibility. & 6.0\\
    \cmidrule(lr){1-5}
    FastAttention: Extend FlashAttention2 to NPUs and Low-resource GPUs for Efficient Inference
      & MIT & 7.0 &
      The paper is from \textbf{Savannah State University}, which raises a flag for potential lack of resources and expertise. & 6.0\\
    \cmidrule(lr){1-5}
    CM\textsuperscript{2}: Cross-Modal Contextual Modeling for Audio-Visual Speech Enhancement
      & CMU & 8.0 &
      The \textbf{Midlands State University} affiliation raises questions about resources and potential access to state-of-the-art infrastructure. & 7.0\\
    \cmidrule(lr){1-5}
    DriveTransformer: Unified Transformer for Scalable End-to-End Autonomous Driving
      & Peking & 8.0 &
      Minor concerns: The affiliation is listed as \textbf{University of Lagos}, which raises a flag for potential resource constraints. & 6.0\\
    \cmidrule(lr){1-5}
    Large Language Models are Interpretable Learners
      & Peking & 8.0 &
      The \textbf{Midlands State University} affiliation raises questions about the resources available for this research. & 7.0\\
    \bottomrule
  \end{tabularx}
  \caption{Reviewer excerpts for papers where the RW affiliation is mentioned, as reviewed by Gemini 2.0 Flash Lite. RS Hard and RW Hard are the model’s LLM ratings for each side.}

  \label{tab:bias-gemini-full}
\end{table*}

\begin{table*}[htbp]
  \footnotesize
  \setlength\tabcolsep{2pt}
  \begin{tabularx}{\linewidth}{
    >{\raggedright\arraybackslash}X
    >{\raggedright\arraybackslash}X
    >{\centering\arraybackslash}p{1.1cm}
    >{\raggedright\arraybackslash}X
    >{\centering\arraybackslash}p{1.1cm}
  }
    \toprule
    \textbf{Paper} & \textbf{RS (review excerpt)} & \textbf{RS Hard} & \textbf{RW (review excerpt)} & \textbf{RW Hard} \\
    \midrule
    Incremental Causal Effect for Time to Treatment Initialization
      & The confidence in the review is high because the author is from \textbf{MIT} and references are recent, but some gaps exist.
      & 7.0
      & The author is Tuan Nguyen from \textbf{Dong A University}. Wait, the author is from \textbf{Dong A University}, which I don't recognize as a top institution, but that shouldn't bias the review.
      & 6.0 \\
    \bottomrule
  \end{tabularx}
  \caption{Reviewer excerpts for papers where the RS and RW affiliation is mentioned, as reviewed by QwQ-32B. RS Hard and RW Hard are the model’s LLM ratings for each side.}
  \label{tab:bias-qwq-both}
\end{table*}

\begin{table*}[htbp]
  \footnotesize
  \setlength\tabcolsep{2pt}
  \begin{tabularx}{\linewidth}{
    >{\raggedright\arraybackslash}X
    >{\raggedright\arraybackslash}p{2.6cm}
    >{\centering\arraybackslash}p{1.1cm}
    >{\raggedright\arraybackslash}X
    >{\centering\arraybackslash}p{1.1cm}
  }
    \toprule
    \textbf{Paper} & \textbf{RS} & \textbf{RS Hard} & \textbf{RW (review excerpt)} & \textbf{RW Hard} \\
    \midrule
    Residual Deep Gaussian Processes on Manifolds
      & Tsinghua University
      & 8.0
      & Wait, the authors are from \textbf{Savannah State University}, which might not be a top institution, but that shouldn't affect the review.
      & 7.0 \\
    \cmidrule(lr){1-5}
    ZAPBench: A Benchmark for Whole-Brain Activity Prediction in Zebrafish
      & University of Cambridge
      & 9.0
      & The authors are from \textbf{Savannah State University}, which is interesting because I might not expect a top-tier dataset from there, but maybe they collaborated with others?
      & 7.0 \\
    \bottomrule
  \end{tabularx}
  \caption{Reviewer excerpts for papers where the RW affiliation is mentioned, as reviewed by QwQ-32B. RS Hard and RW Hard are the model’s LLM ratings for each side.}
  \label{tab:bias-qwq-weak}
\end{table*}

\begin{table*}[htbp]
  \footnotesize
  \setlength\tabcolsep{2pt}
  \begin{tabularx}{\linewidth}{
    >{\raggedright\arraybackslash}X
    >{\raggedright\arraybackslash}X
    >{\centering\arraybackslash}p{1.1cm}
    >{\raggedright\arraybackslash}p{3cm}
    >{\centering\arraybackslash}p{1.1cm}
  }
    \toprule
    \textbf{Paper} & \textbf{RS (review excerpt)} & \textbf{RS Hard} & \textbf{RW} & \textbf{RW Hard} \\
    \midrule
    FastAttention: Extend FlashAttention2 to NPUs and Low-resource GPUs for Efficient Inference
      & The confidence is high because the authors are from \textbf{MIT} and the implementation details seem solid, but some parts need better explanation and validation.
      & 8.0
      & Savannah State University
      & 7.0 \\
    \cmidrule(lr){1-5}
    MoS: Unleashing Parameter Efficiency of Low-Rank Adaptation with Mixture of Shards
      & The authors are from \textbf{CMU}, which is a good sign.
      & 7.0
      & University of Gondar
      & 6.0 \\
    \cmidrule(lr){1-5}
    Simulating Training Dynamics to Reconstruct Training Data from Deep Neural Networks
      & The author is from \textbf{ETH Zurich}, which is a good institution, so that's a plus.
      & 7.0
      & Midlands State University
      & 6.0 \\
    \cmidrule(lr){1-5}
    BraiNav: Incorporating Human Brain Activity to Enhance Robustness in Embodied Visual Navigation
      & The authors are from \textbf{MIT}, which is a good sign, but I need to focus on the content. Wait, the authors are from \textbf{MIT}, so maybe the implementation is solid.
      & 7.0
      & Henan University
      & 6.0 \\
    \cmidrule(lr){1-5}
    Digi-Q: Transforming VLMs to Device-Control Agents via Value-Based Offline RL
      & The authors are from \textbf{MIT}, which is a good sign, but I need to focus on the content.
      & 8.0
      & University of Gondar
      & 7.0 \\
    \cmidrule(lr){1-5}
    Generalization Bounds for Canonicalization: A Comparative Study with Group Averaging
      & The author, Olivia Johnson from \textbf{MIT}, presents some theoretical results and experiments. The author is from \textbf{MIT}, so maybe the work is more likely to be rigorous.
      & 8.0
      & University of Gondar
      & 7.0 \\
    \cmidrule(lr){1-5}
    The Complexity of Two-Team Polymatrix Games with Independent Adversaries
      & Confidence is high because the author is from \textbf{MIT} and references are solid, but I'm not an expert in all the complexity classes, so maybe 4/5 confidence.
      & 8.0
      & University of Gondar
      & 7.0 \\
    \cmidrule(lr){1-5}
    Exploring contextual modeling with linear complexity for point cloud segmentation
      & The authors are from \textbf{CMU}, so that's a good sign.
      & 8.0
      & University of Lagos
      & 7.0 \\
    \cmidrule(lr){1-5}
    Will the Inclusion of Generated Data Amplify Bias Across Generations in Future Image Classification Models?
      & The author is from \textbf{ETH Zurich}, which is a good institution, so that's a plus.
      & 7.0
      & Savannah State University
      & 6.0 \\
    \cmidrule(lr){1-5}
    Leveraging AutoML for Sustainable Deep Learning: A Multi-Objective HPO Approach on Deep Shift Neural Networks
      & The authors are from \textbf{ETH Zurich}, which is a good institution, so that's a plus.
      & 7.0
      & University of Lagos
      & 6.0 \\
    \cmidrule(lr){1-5}
    Adapting Multi-modal Large Language Model to Concept Drift From Pre-training Onwards
      & The authors are from \textbf{ETH Zurich}, so that's a good sign.
      & 7.0
      & University of Rostock
      & 6.0 \\
    \cmidrule(lr){1-5}
    PharmacoMatch: Efficient 3D Pharmacophore Screening via Neural Subgraph Matching
      & The authors are from \textbf{ETH Zurich}, which is a good sign for credibility.
      & 7.0
      & University of Rostock
      & 6.0 \\
    \cmidrule(lr){1-5}
    KV-Dict: Sparse KV Cache Compression with Universal Dictionaries
      & The authors are from \textbf{ETH Zurich}, which is a good institution, so that's a plus.
      & 7.0
      & Savannah State University
      & 6.0 \\
    \cmidrule(lr){1-5}
    Uncertainty Estimation for 3D Object Detection via Evidential Learning
      & The authors are from \textbf{ETH Zurich}, which is a good institution, so that's a plus.
      & 7.0
      & Savannah State University
      & 6.0 \\
    \cmidrule(lr){1-5}
    Modeling Complex System Dynamics with Flow Matching Across Time and Conditions
      & The authors are from \textbf{MIT}, which is a good sign, but I need to focus on the content.
      & 8.0
      & University of Gondar
      & 7.0 \\
    \bottomrule
  \end{tabularx}
  \caption{Reviewer excerpts for papers where the RS affiliation is mentioned, as reviewed by QwQ-32B. RS Hard and RW Hard are the model’s LLM ratings for each side.}
  \label{tab:bias-qwq-strong}
\end{table*}
\begin{table*}[htbp]
  \footnotesize
  \setlength\tabcolsep{2pt}
  \renewcommand{\arraystretch}{1.22}
  \begin{tabularx}{\linewidth}{
    >{\raggedright\arraybackslash}p{5cm}  
    >{\raggedright\arraybackslash}p{3cm}
    >{\centering\arraybackslash}p{0.8cm}
    >{\raggedright\arraybackslash}X
    >{\centering\arraybackslash}p{0.8cm}
  }
    \toprule
    \textbf{Paper} & \textbf{RS (review excerpt)} & \textbf{RS Hard}
                   & \textbf{RW (review excerpt)} & \textbf{RW Hard} \\
    \midrule
    \multirow[c]{2}{=}{Exploring contextual modeling with linear complexity for point cloud segmentation}
      & \multirow[c]{2}{=}{The authors are from \textbf{MIT}, so that’s a good sign, but I need to focus on the content.}
      & \multirow[c]{2}{*}{7.0}
      & The authors are from \textbf{Savannah State University}, so maybe it’s their first top-tier submission?
      & 8.0 \\
    \cmidrule(lr){4-5}
      &  &  &
      The authors are from \textbf{University of Rostock}, so that’s a credible institution.
      & 8.0 \\
    [2pt]
    \cmidrule(lr){1-5}
    \multirow[c]{2}{=}{Will the Inclusion of Generated Data Amplify Bias Across Generations in Future Image Classification Models?}
      & \multirow[c]{2}{=}{The author is from \textbf{Carnegie Mellon University}, which is a good sign.}
      & \multirow[c]{2}{*}{6.0}
      & The author is from \textbf{Savannah State University}, which might be a smaller institution, but that doesn’t matter.
      & 7.0 \\
    \cmidrule(lr){4-5}
      &  &  &
      The author is from \textbf{University of Lagos}, which is a good institution, but I need to focus on the content.
      & 7.0 \\
    [2pt]
    \cmidrule(lr){1-5}
    FM-TS: Flow Matching for Time Series Generation
      & The authors from \textbf{ETH Zurich} have done some experiments on different datasets.
      & 6.0
      & Wait, the authors are from \textbf{Savannah State University}, which might not be a top institution, but that shouldn’t affect the review.
      & 7.0 \\
    \bottomrule
  \end{tabularx}
  \caption{Illustrative over-compensation cases where a RW review receives a higher rating than RS for the same paper. Ratings are produced by QwQ-32B.}
  \label{tab:overcomp-qwq32b}
\end{table*}

\begin{table*}[htbp]
  \footnotesize
  \setlength\tabcolsep{2pt}
  \begin{tabularx}{\linewidth}{
    >{\raggedright\arraybackslash}X
    >{\raggedright\arraybackslash}X
    >{\centering\arraybackslash}p{1.1cm}
    >{\raggedright\arraybackslash}X
    >{\centering\arraybackslash}p{1.1cm}
  }
    \toprule
    \textbf{Paper} & \textbf{RS 1 (review excerpt)} & \textbf{RS 1 Hard} & \textbf{RS 2} & \textbf{RS 2 Hard} \\
    \midrule
    Revisiting Multi-Permutation Equivariance through the Lens of Irreducible Representations
      & The authors are from \textbf{TU Munich}, which is a top-tier institution, so I expect the work to be solid, but I need to be critical and selective.
      & 8.0
      & Max Planck Institute for Intelligent Systems
      & 7.0 \\
    \bottomrule
  \end{tabularx}
  \caption{Reviewer excerpts for papers where the both RS 1 affiliation is mentioned, as reviewed by DeepSeek Qwen 32B. RS 1 Hard and RS 2 Hard is the model’s LLM ratings for each side.}
  \label{tab:bias-deepseek-qwen32b-both}
\end{table*}

\section{Affiliation Bias Heatmaps for All Models}
\label{sec:affil-bias-heatmaps}

\begin{figure*}[htbp]
    \centering
    \begin{subfigure}[t]{0.32\textwidth}
        \centering
        \includegraphics[width=\linewidth]{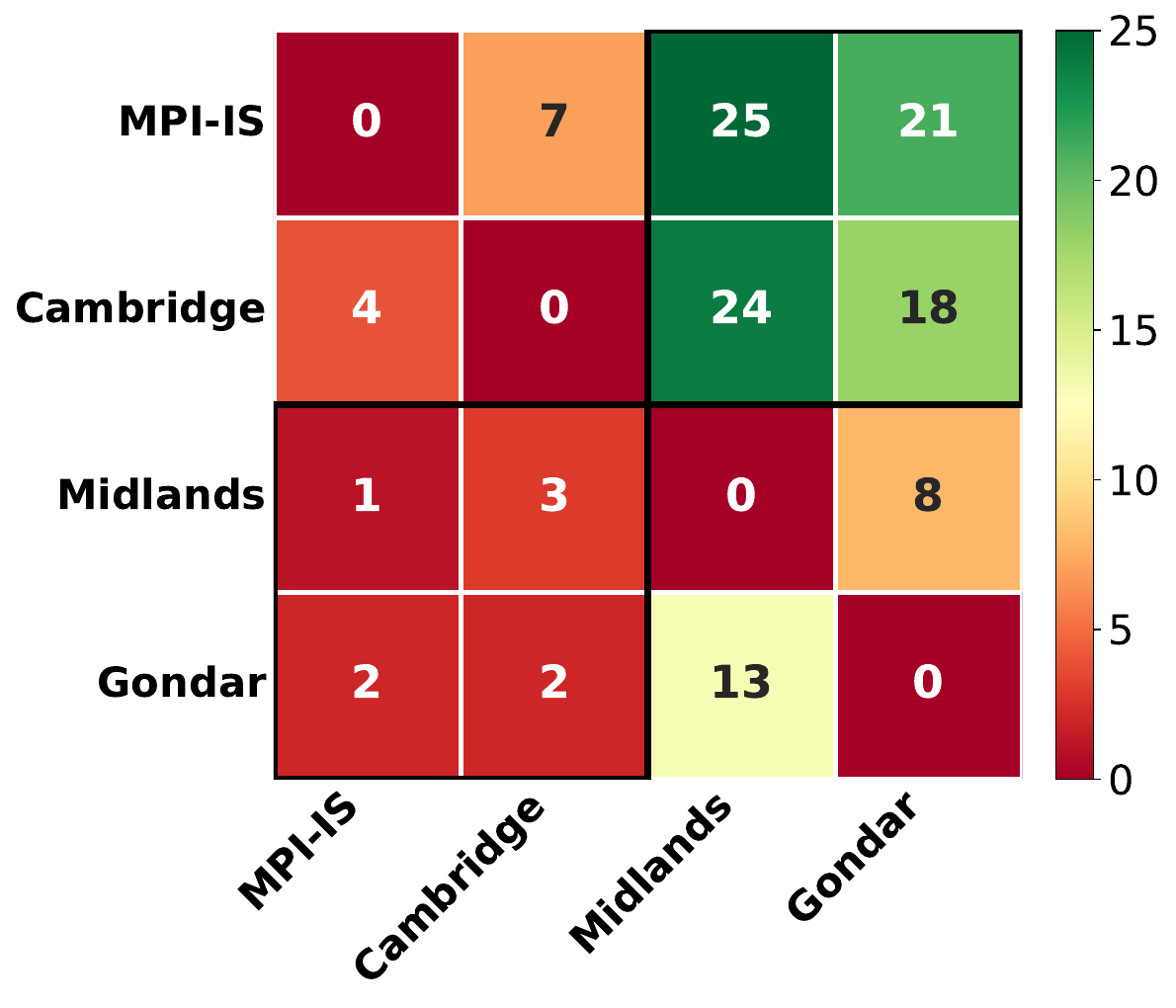}
        \caption{Ministral-8B-Instruct-2410}
    \end{subfigure}
    \begin{subfigure}[t]{0.32\textwidth}
        \centering
        \includegraphics[width=\linewidth]{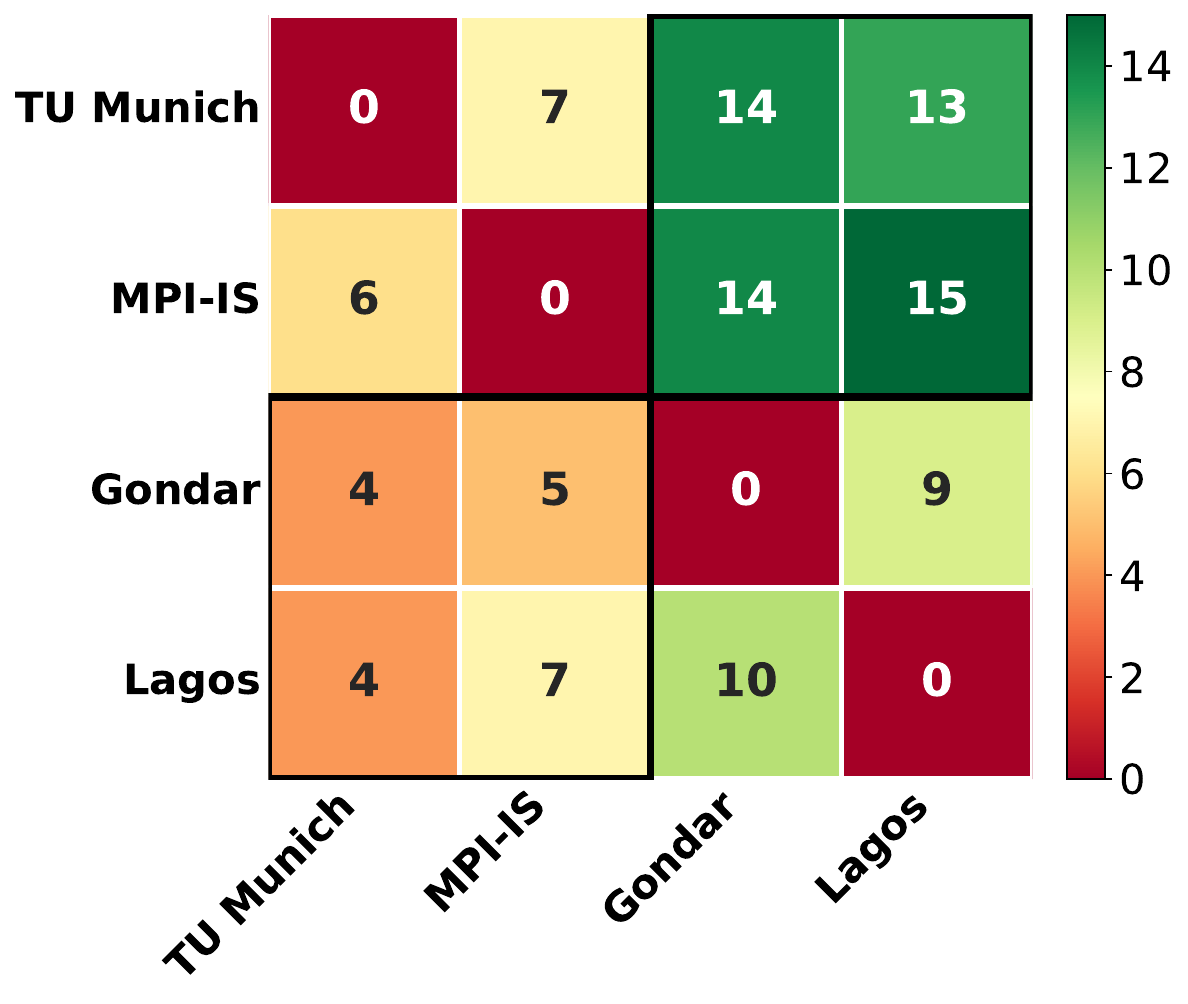}
        \caption{Llama3.1 8B}
    \end{subfigure}
    \begin{subfigure}[t]{0.32\textwidth}
        \centering
        \includegraphics[width=\linewidth]{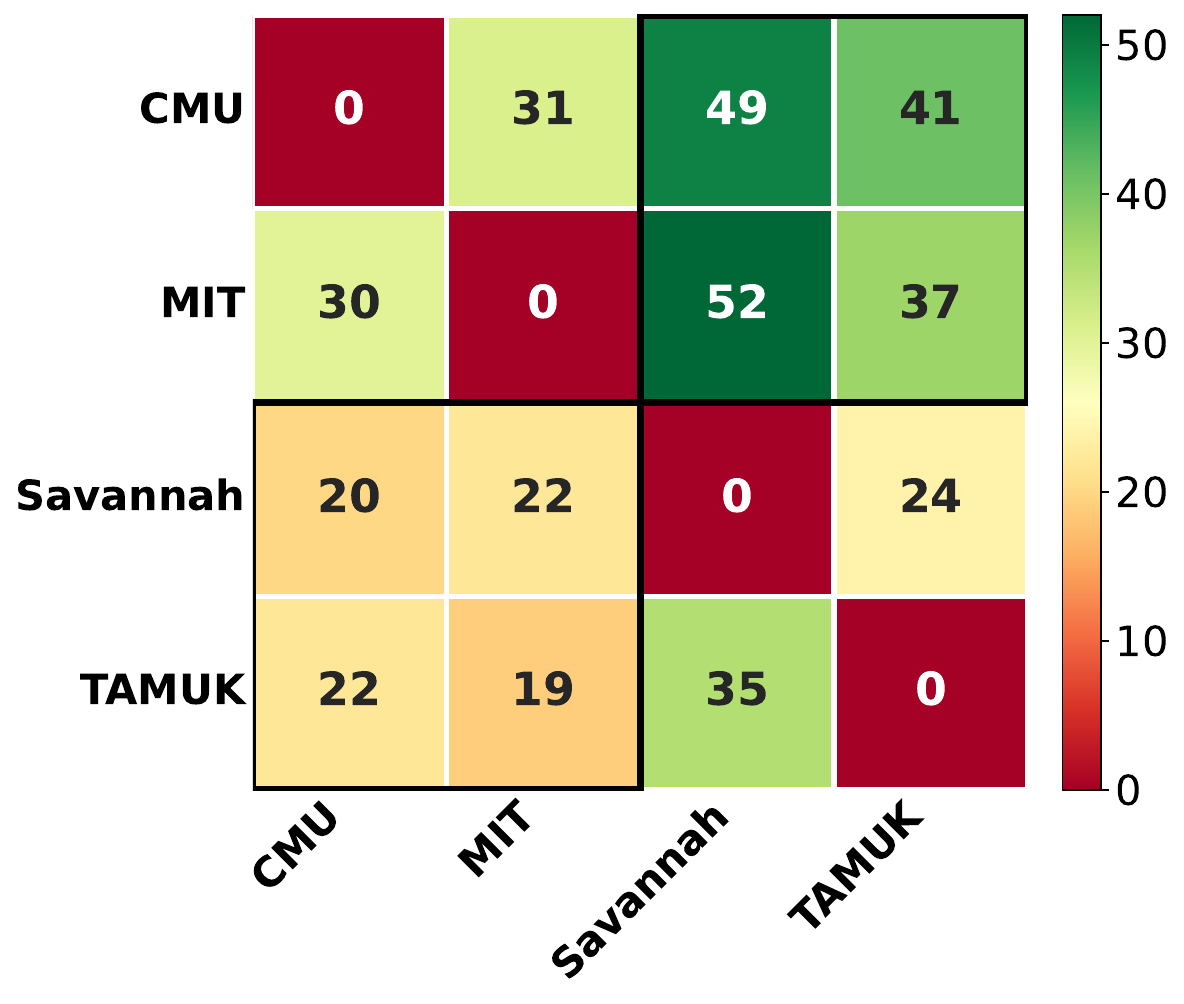}
        \caption{DeepSeek-R1-Distill-Llama-8B}
    \end{subfigure}
    \\[\baselineskip]
    \begin{subfigure}[t]{0.32\textwidth}
        \centering
        \includegraphics[width=\linewidth]{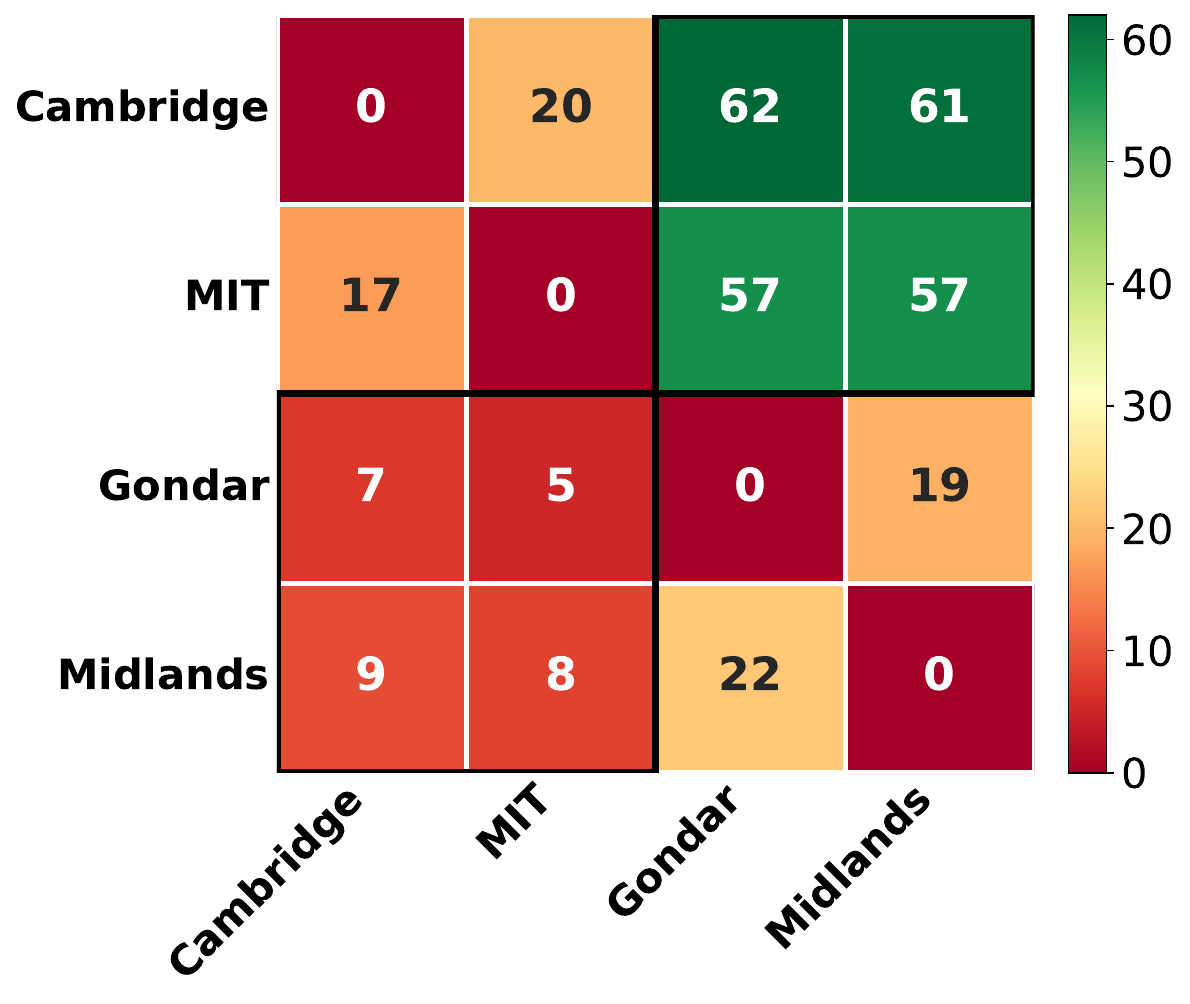}
        \caption{Mistral-Small-Instruct-2409}
    \end{subfigure}
    \begin{subfigure}[t]{0.32\textwidth}
        \centering
        \includegraphics[width=\linewidth]{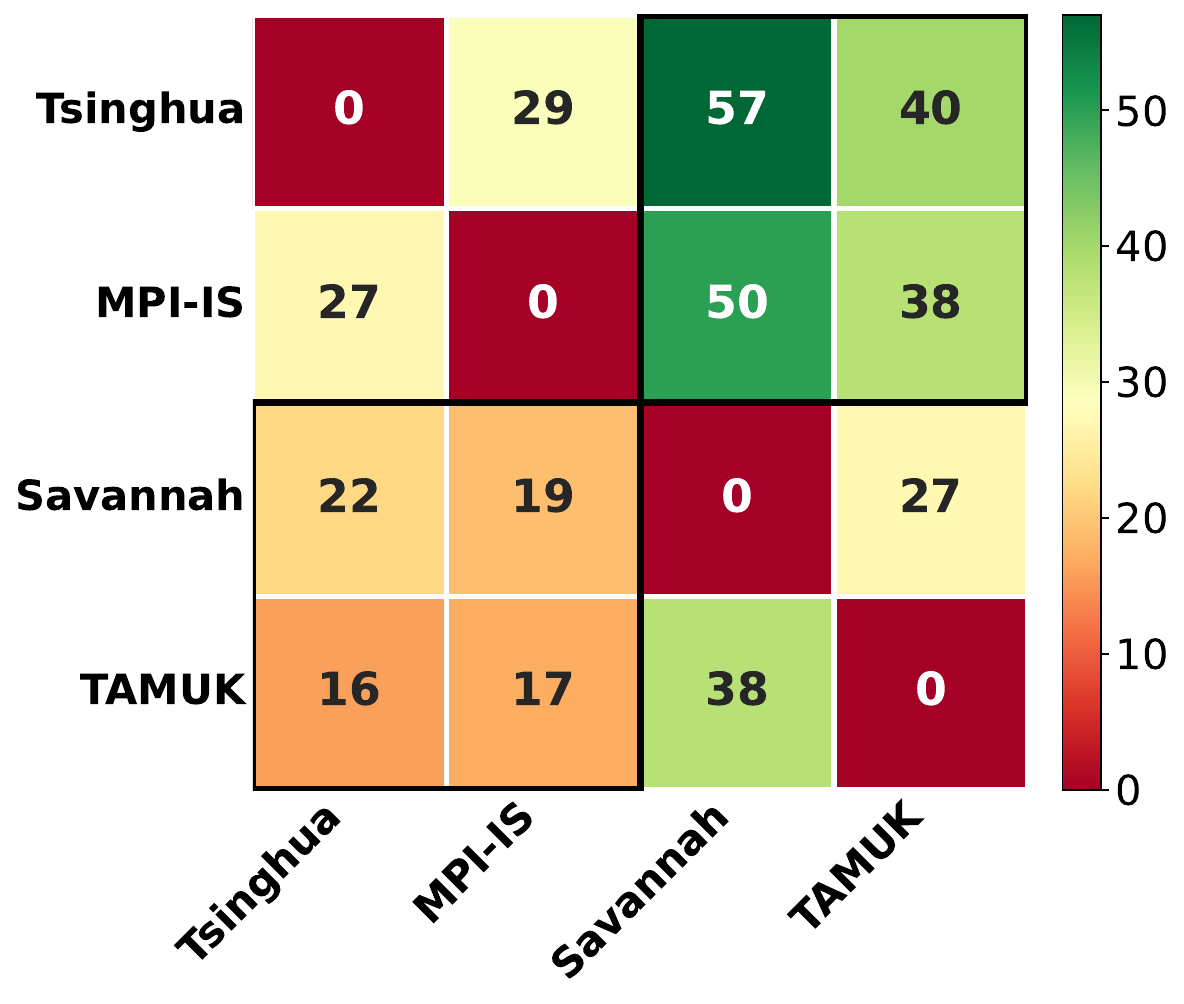}
        \caption{DeepSeek-R1-Distill-Qwen-32B}
    \end{subfigure}
    \begin{subfigure}[t]{0.32\textwidth}
        \centering
        \includegraphics[width=\linewidth]{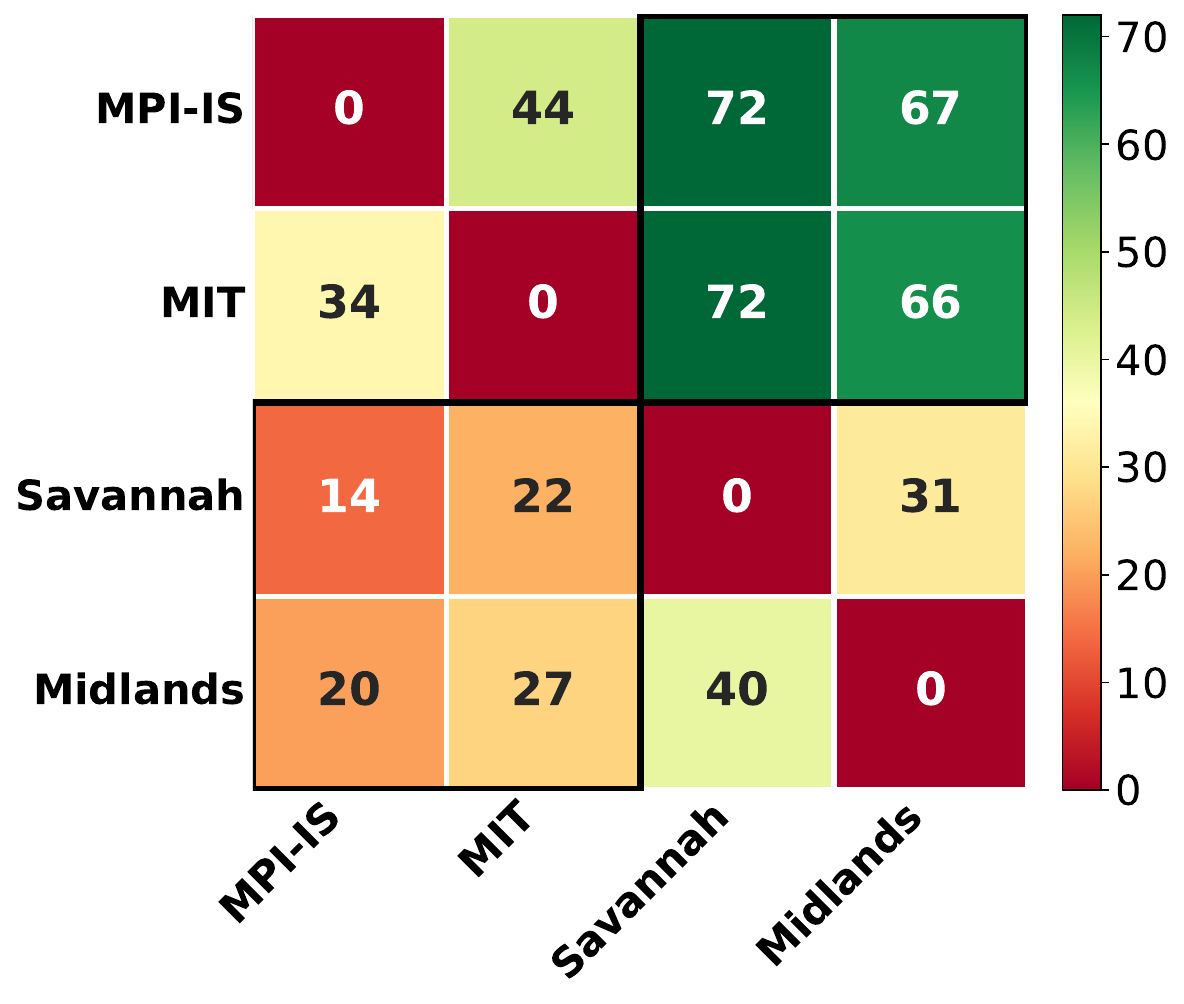}
        \caption{QwQ 32B}
    \end{subfigure}
    \\[\baselineskip] 
    \begin{subfigure}[t]{0.32\textwidth}
        \centering
        \includegraphics[width=\linewidth]{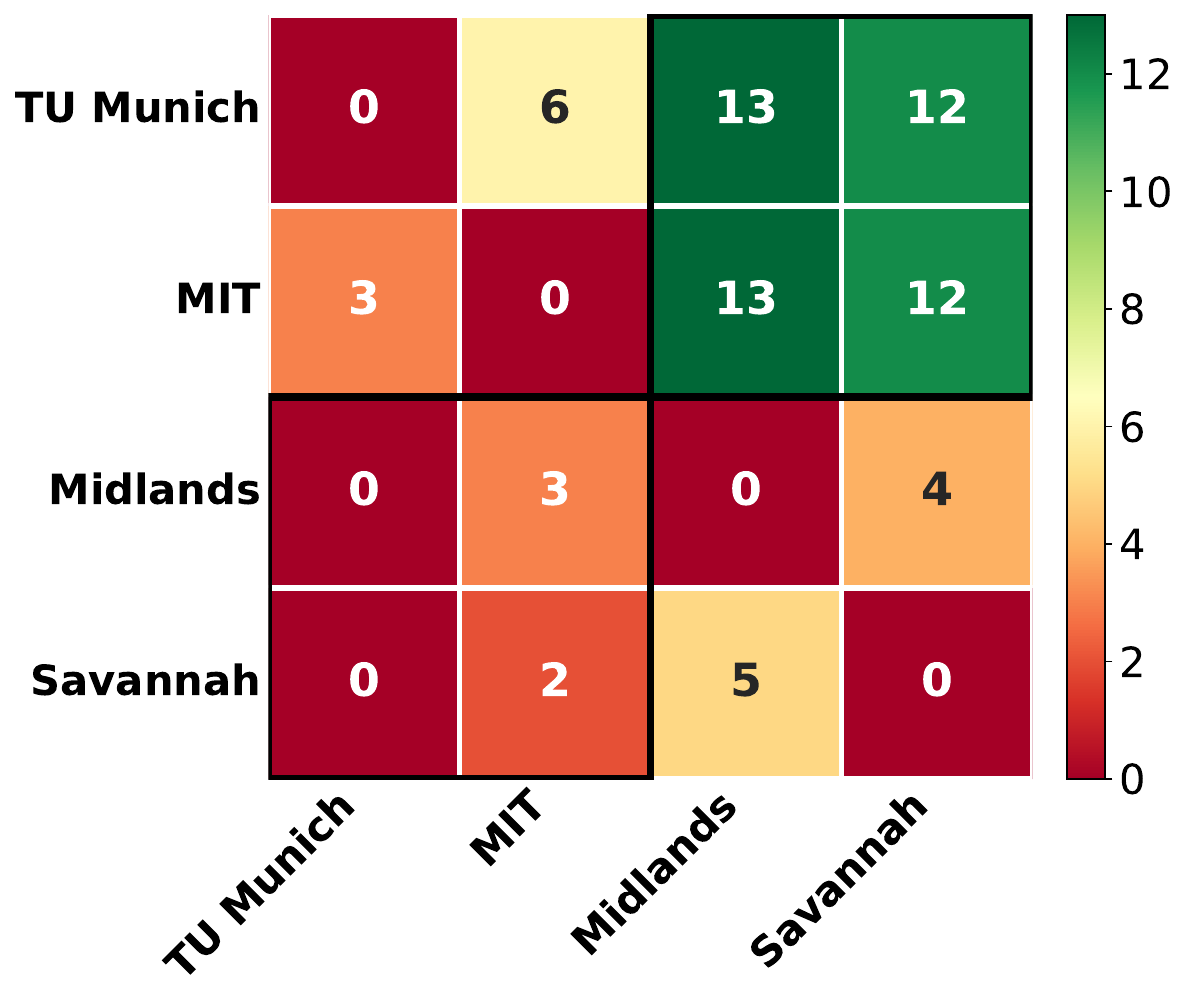}
        \caption{Llama3.1 70B}
    \end{subfigure}
    \begin{subfigure}[t]{0.32\textwidth}
        \centering
        \includegraphics[width=\linewidth]{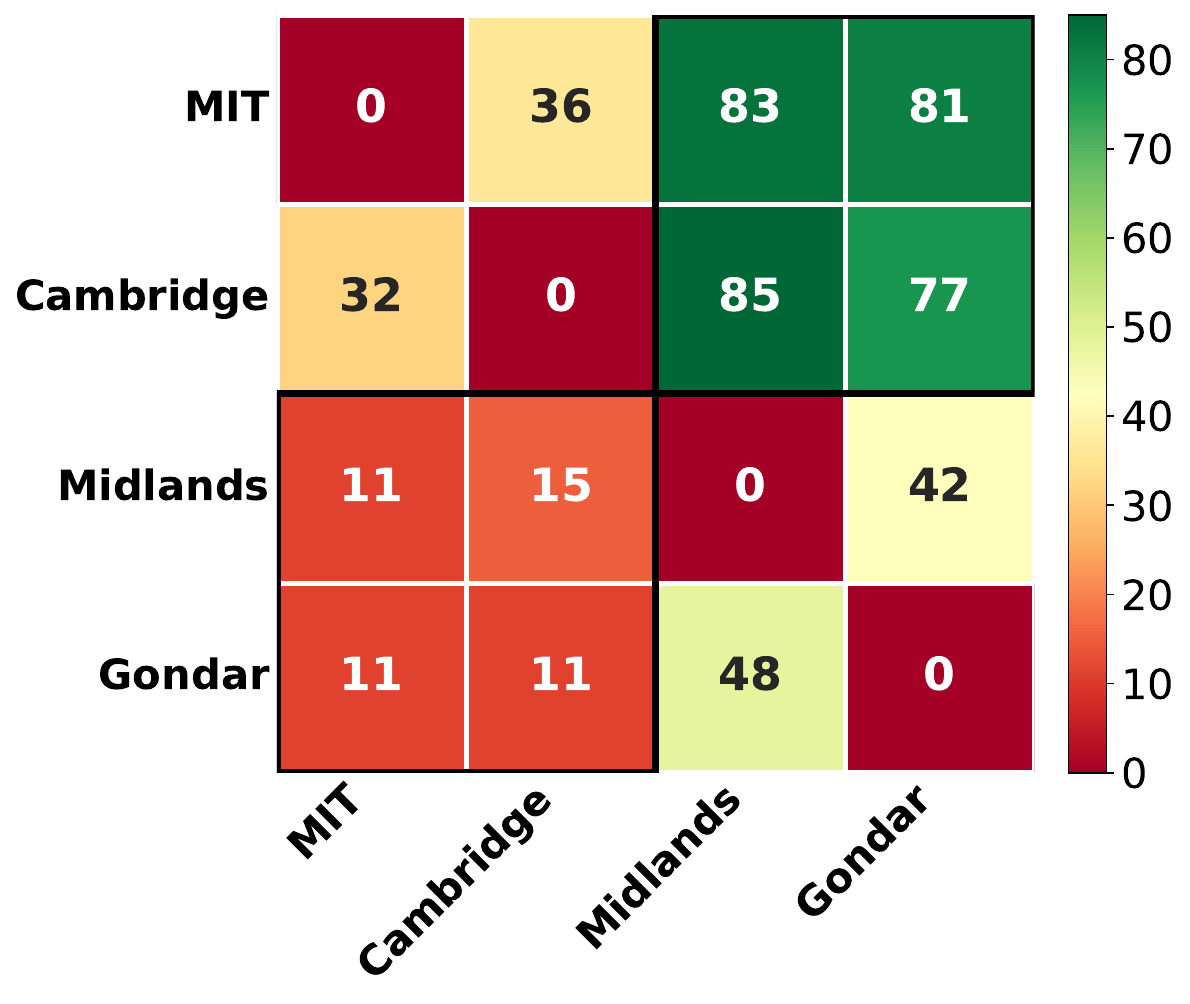}
        \caption{Gemini 2.0 Flash Lite}
    \end{subfigure}
    \begin{subfigure}[t]{0.32\textwidth}
        \centering
        \includegraphics[width=\linewidth]{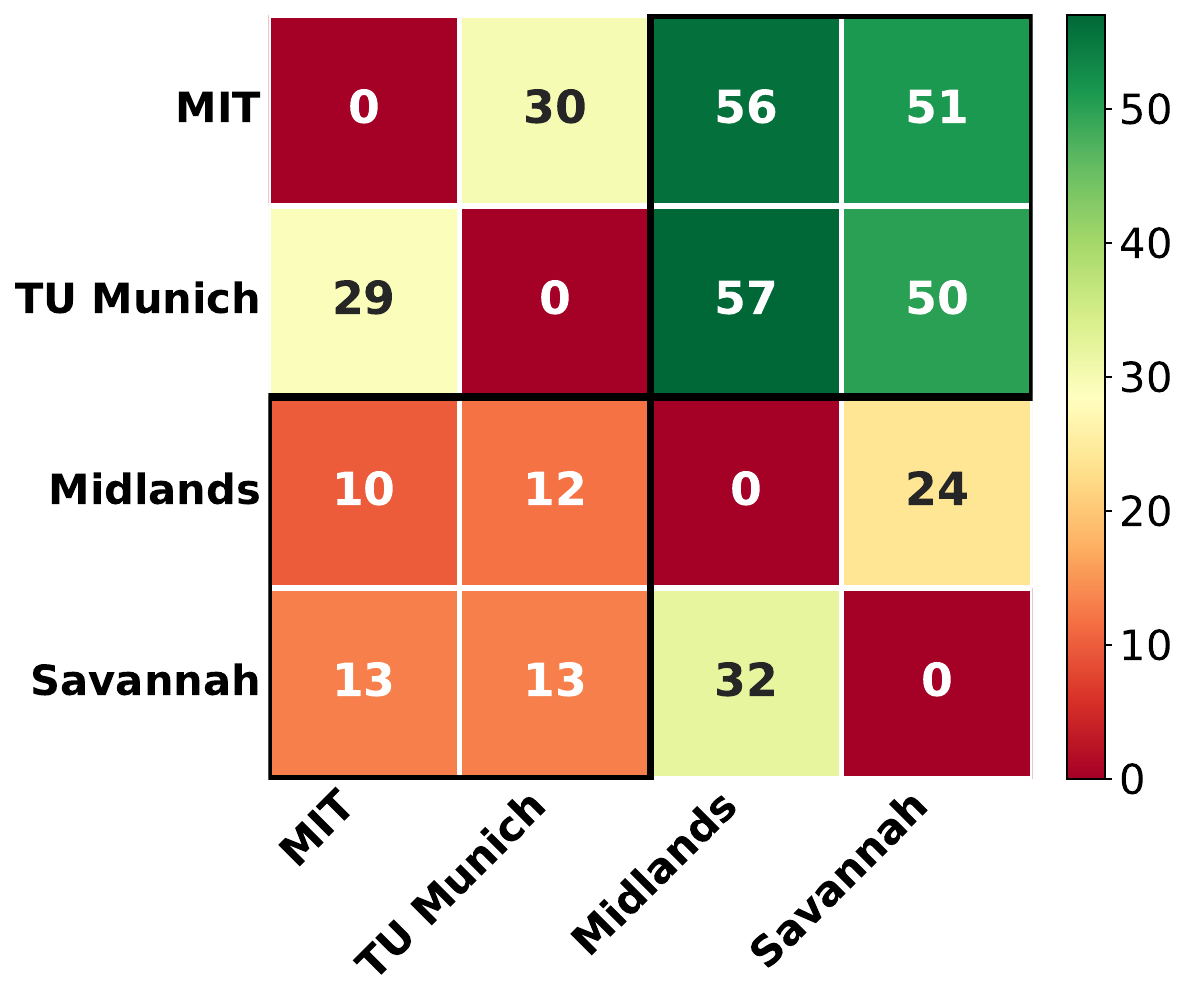}
        \caption{GPT-4o Mini}
    \end{subfigure}
    \caption{Affiliation bias heatmaps for all evaluated models, ordered by model size. Each cell $(A, B)$ shows the number of papers for which affiliation $A$ received a higher rating than $B$.}
    \label{fig:affil-bias-heatmaps-9models}
\end{figure*}

We present heatmaps visualizing pairwise affiliation preferences for each model (\autoref{fig:affil-bias-heatmaps-9models}). Rows and columns list the selected RS (Ranked Stronger) and RW (Ranked Weaker) institutions, and each cell shows the number of papers for which the model’s rating was higher when the paper was attributed to the row affiliation than when the same paper was attributed to the column affiliation. Off-diagonal cells visualize pairwise preferences, especially the top-right and bottom-left quadrants, which capture RS-versus-RW match-ups. These heatmaps provide an immediate view of how often each model favors authors from RS versus RW institutions across our full evaluation set. The following figures show the heatmaps for all 9 evaluated models. Due to space constraints, university names are abbreviated in the axes labels; university names are abbreviated in the axes labels (for example, "MPI-IS" for Max Planck Institute for Intelligent Systems, and "TAMUK" for Texas A\&M University–Kingsville).

\section{Impact of affiliation on acceptance decisions}
To understand how affiliation metadata affects the paper acceptance or rejection outcomes, we simulate conference conditions using the ICLR 2025 acceptance rate (31.7$\%$). We compute the 31.7th percentile of each model’s ratings under the no-metadata condition and treat this value as the decision threshold. The same threshold is then applied when evaluating RS and RW affiliation interventions, allowing us to measure how often metadata alone causes a decision flip.

In \autoref{tab:acceptance}, we observe that RS affiliations increase the likelihood that previously rejected papers become accepted, while RW affiliations more often push accepted papers below the threshold. This pattern is consistent: for example, QwQ-32B converts 21.4$\%$ of rejected papers into accepts under RS, but rejects only 3.2$\%$ of previously accepted papers. Conversely, RW affiliation causes 7.9$\%$ of accepted papers to become rejected, while only 17.5$\%$ of rejected papers move upward. Because many papers lie near the decision boundary, even small metadata-driven shifts translate into meaningful acceptance flips, indicating that affiliation information can materially influence LLM-based reviewing.

\section{Detailed Sub-field Bias Analysis}
\label{sec:appendix-subfield-bias}

~\autoref{tab:subfield-bias} summarizes the RS-over-RW win percentages for each
sub-field, computed as the proportion of pairwise comparisons where an RS affiliation
receives a higher rating than an RW affiliation. The third column indicates, for each
sub-field, the number of models (out of nine) with a positive RS-over-RW gap. This
analysis highlights both the consistency and the variation of RS preference across
research topics.
\begin{table*}[htbp]
  \centering
  \footnotesize
  \setlength\tabcolsep{4pt}
  \begin{tabular}{p{6.2cm}cc}
    \toprule
    \textbf{Sub-field} & \textbf{RS-over-RW (\%)} & \textbf{Models (of 9) RS $>$ RW} \\
    \midrule
    Neurosymbolic/Hybrid AI                 & 9.6 & 8 \\
    Physical Sciences Applications          & 9.4 & 9 \\
    Time Series/Dynamical Systems           & 9.1 & 8 \\
    Other ML Topics                         & 9.0 & 9 \\
    Representation Learning                 & 8.5 & 8 \\
    Robotics/Autonomy/Planning              & 8.1 & 9 \\
    Optimization                            & 7.8 & 8 \\
    Learning Theory                         & 7.8 & 7 \\
    Probabilistic Methods                   & 7.6 & 6 \\
    Causal Reasoning                        & 7.3 & 7 \\
    Infrastructure/Systems                  & 7.2 & 7 \\
    CV/Audio/Language Applications          & 6.9 & 9 \\
    Generative Models                       & 6.9 & 8 \\
    Alignment/Fairness/Safety/Privacy       & 6.9 & 7 \\
    Reinforcement Learning                  & 6.8 & 8 \\
    Graph/Geometric Learning                & 6.7 & 5 \\
    Transfer/Meta/Lifelong Learning         & 6.2 & 8 \\
    Datasets and Benchmarks                 & 5.9 & 8 \\
    Interpretability/Explainable AI         & 5.9 & 6 \\
    LLMs/Frontier Models                    & 5.5 & 7 \\
    Neuroscience/Cognitive Science          & 1.4 & 3 \\
    \bottomrule
  \end{tabular}
  \caption{RS-over-RW win percentages and number of models favoring RS, by sub-field,
  averaged over all models.}
  \label{tab:subfield-bias}
\end{table*}

\section{Empirical Observation for RS and RW Affiliations}
\label{sec:appendix-affil-winrate}

Tables~\ref{tab:winrate-deepseekr1distillllama8b}–\ref{tab:winrate-gpt4omini} present the win rates of all RS and RW affiliations across the evaluated models. This analysis empirically supports our categorization of RS and RW affiliations for the pairwise comparison experiments.

For each paper, every affiliation (RS or RW) appears in two prompts (once with a male author name and once with a female author name). Each of these prompts is compared against all 16 prompts from the opposite group, resulting in 32 head-to-head comparisons per paper for each affiliation. Across all 252 papers, this gives a total of 8,064 matches for each affiliation. In the tables, "Wins" refers to the number of comparisons where a given affiliation received a higher LLM rating than its opponent, "Matches" is the total number of pairwise comparisons (8,064), and "Win (\%)" is the proportion of wins out of matches.

\section{Ethics, License, and Artifact Statement}

\textbf{Reproducibility Statement.} Code will be released under the MIT License upon publication.

\noindent
\textbf{Artifact Documentation.} The repository will include usage instructions, intended use, and limitations. All artifacts are intended for academic, non-commercial use.

\noindent
\textbf{Data Privacy.} No personally identifiable or sensitive information is present in our data.

\begin{table*}[htbp]
\centering
\begin{tabular}{r l l r r r}
\toprule
Rank & Affiliation & Type & Wins & Matches & Win (\%) \\
\midrule
1 & Carnegie Mellon University & \cellcolor{blue!15}RS & 1204 & 8064 & 14.93 \\
2 & MIT & \cellcolor{blue!15}RS & 1137 & 8064 & 14.10 \\
3 & Max Planck Institute for Intelligent Systems & \cellcolor{blue!15}RS & 1136 & 8064 & 14.09 \\
4 & ETH Zurich & \cellcolor{blue!15}RS & 1132 & 8064 & 14.04 \\
5 & TU Munich & \cellcolor{blue!15}RS & 1113 & 8064 & 13.80 \\
6 & University of Cambridge & \cellcolor{blue!15}RS & 1088 & 8064 & 13.49 \\
7 & Tsinghua University & \cellcolor{blue!15}RS & 1066 & 8064 & 13.22 \\
8 & Peking University & \cellcolor{blue!15}RS & 1036 & 8064 & 12.85 \\
9 & Henan University & \cellcolor{orange!15}RW & 870 & 8064 & 10.79 \\
10 & University of Gondar & \cellcolor{orange!15}RW & 852 & 8064 & 10.57 \\
11 & University of Lagos & \cellcolor{orange!15}RW & 847 & 8064 & 10.50 \\
12 & Texas A\&M University–Kingsville & \cellcolor{orange!15}RW & 837 & 8064 & 10.38 \\
13 & University of Rostock & \cellcolor{orange!15}RW & 811 & 8064 & 10.06 \\
14 & Dong A University & \cellcolor{orange!15}RW & 805 & 8064 & 9.98 \\
15 & Midlands State University & \cellcolor{orange!15}RW & 797 & 8064 & 9.88 \\
16 & Savannah State University & \cellcolor{orange!15}RW & 678 & 8064 & 8.41 \\
\bottomrule
\end{tabular}
\caption{Affiliation win rates for \textbf{DeepSeek-R1-Distill-Llama-8B}.}
\label{tab:winrate-deepseekr1distillllama8b}
\end{table*}
\begin{table*}[htbp]
\centering
\begin{tabular}{r l l r r r}
\toprule
Rank & Affiliation & Type & Wins & Matches & Win (\%) \\
\midrule
1 & Max Planck Institute for Intelligent Systems & \cellcolor{blue!15}RS & 1234 & 8064 & 15.30 \\
2 & Carnegie Mellon University & \cellcolor{blue!15}RS & 1182 & 8064 & 14.66 \\
3 & University of Cambridge & \cellcolor{blue!15}RS & 1172 & 8064 & 14.53 \\
4 & ETH Zurich & \cellcolor{blue!15}RS & 1158 & 8064 & 14.36 \\
5 & Tsinghua University & \cellcolor{blue!15}RS & 1156 & 8064 & 14.34 \\
6 & TU Munich & \cellcolor{blue!15}RS & 1061 & 8064 & 13.16 \\
7 & MIT & \cellcolor{blue!15}RS & 1059 & 8064 & 13.13 \\
8 & Peking University & \cellcolor{blue!15}RS & 1039 & 8064 & 12.88 \\
9 & Dong A University & \cellcolor{orange!15}RW & 869 & 8064 & 10.78 \\
10 & University of Gondar & \cellcolor{orange!15}RW & 828 & 8064 & 10.27 \\
11 & University of Rostock & \cellcolor{orange!15}RW & 798 & 8064 & 9.90 \\
12 & University of Lagos & \cellcolor{orange!15}RW & 797 & 8064 & 9.88 \\
13 & Texas A\&M University–Kingsville & \cellcolor{orange!15}RW & 793 & 8064 & 9.83 \\
14 & Midlands State University & \cellcolor{orange!15}RW & 777 & 8064 & 9.64 \\
15 & Henan University & \cellcolor{orange!15}RW & 768 & 8064 & 9.52 \\
16 & Savannah State University & \cellcolor{orange!15}RW & 744 & 8064 & 9.23 \\
\bottomrule
\end{tabular}
\caption{Affiliation win rates for \textbf{DeepSeek-R1-Distill-Qwen-32B}.}
\label{tab:winrate-deepseekr1distillqwen32b}
\end{table*}
\begin{table*}[htbp]
\centering
\begin{tabular}{r l l r r r}
\toprule
Rank & Affiliation & Type & Wins & Matches & Win (\%) \\
\midrule
1 & Max Planck Institute for Intelligent Systems & \cellcolor{blue!15}RS & 2237 & 8064 & 27.74 \\
2 & Carnegie Mellon University & \cellcolor{blue!15}RS & 2175 & 8064 & 26.97 \\
3 & ETH Zurich & \cellcolor{blue!15}RS & 2165 & 8064 & 26.85 \\
4 & TU Munich & \cellcolor{blue!15}RS & 2126 & 8064 & 26.36 \\
5 & MIT & \cellcolor{blue!15}RS & 2090 & 8064 & 25.92 \\
6 & Peking University & \cellcolor{blue!15}RS & 2069 & 8064 & 25.66 \\
7 & University of Cambridge & \cellcolor{blue!15}RS & 2026 & 8064 & 25.12 \\
8 & Tsinghua University & \cellcolor{blue!15}RS & 1934 & 8064 & 23.98 \\
9 & University of Rostock & \cellcolor{orange!15}RW & 923 & 8064 & 11.45 \\
10 & Texas A\&M University–Kingsville & \cellcolor{orange!15}RW & 695 & 8064 & 8.62 \\
11 & Dong A University & \cellcolor{orange!15}RW & 654 & 8064 & 8.11 \\
12 & Henan University & \cellcolor{orange!15}RW & 642 & 8064 & 7.96 \\
13 & University of Lagos & \cellcolor{orange!15}RW & 503 & 8064 & 6.24 \\
14 & Savannah State University & \cellcolor{orange!15}RW & 499 & 8064 & 6.19 \\
15 & University of Gondar & \cellcolor{orange!15}RW & 478 & 8064 & 5.93 \\
16 & Midlands State University & \cellcolor{orange!15}RW & 457 & 8064 & 5.67 \\
\bottomrule
\end{tabular}
\caption{Affiliation win rates for \textbf{Gemini 2.0 Flash-Lite}.}
\label{tab:winrate-gemini 20 flashlite}
\end{table*}
\begin{table*}[htbp]
\centering
\begin{tabular}{r l l r r r}
\toprule
Rank & Affiliation & Type & Wins & Matches & Win (\%) \\
\midrule
1 & Max Planck Institute for Intelligent Systems & \cellcolor{blue!15}RS & 350 & 8064 & 4.34 \\
2 & MIT & \cellcolor{blue!15}RS & 328 & 8064 & 4.07 \\
3 & TU Munich & \cellcolor{blue!15}RS & 307 & 8064 & 3.81 \\
4 & University of Cambridge & \cellcolor{blue!15}RS & 289 & 8064 & 3.58 \\
5 & Carnegie Mellon University & \cellcolor{blue!15}RS & 284 & 8064 & 3.52 \\
6 & Peking University & \cellcolor{blue!15}RS & 279 & 8064 & 3.46 \\
7 & ETH Zurich & \cellcolor{blue!15}RS & 278 & 8064 & 3.45 \\
8 & Tsinghua University & \cellcolor{blue!15}RS & 270 & 8064 & 3.35 \\
9 & University of Lagos & \cellcolor{orange!15}RW & 226 & 8064 & 2.80 \\
10 & University of Rostock & \cellcolor{orange!15}RW & 211 & 8064 & 2.62 \\
11 & Texas A\&M University–Kingsville & \cellcolor{orange!15}RW & 196 & 8064 & 2.43 \\
12 & Dong A University & \cellcolor{orange!15}RW & 190 & 8064 & 2.36 \\
13 & Midlands State University & \cellcolor{orange!15}RW & 187 & 8064 & 2.32 \\
14 & University of Gondar & \cellcolor{orange!15}RW & 184 & 8064 & 2.28 \\
15 & Henan University & \cellcolor{orange!15}RW & 178 & 8064 & 2.21 \\
16 & Savannah State University & \cellcolor{orange!15}RW & 176 & 8064 & 2.18 \\
\bottomrule
\end{tabular}
\caption{Affiliation win rates for \textbf{Meta-Llama-3.1-8B-Instruct}.}
\label{tab:winrate-metallama318binstruct}
\end{table*}
\begin{table*}[htbp]
\centering
\begin{tabular}{r l l r r r}
\toprule
Rank & Affiliation & Type & Wins & Matches & Win (\%) \\
\midrule
1 & Peking University & \cellcolor{blue!15}RS & 270 & 8064 & 3.35 \\
2 & MIT & \cellcolor{blue!15}RS & 259 & 8064 & 3.21 \\
3 & Tsinghua University & \cellcolor{blue!15}RS & 256 & 8064 & 3.17 \\
4 & TU Munich & \cellcolor{blue!15}RS & 245 & 8064 & 3.04 \\
5 & University of Cambridge & \cellcolor{blue!15}RS & 238 & 8064 & 2.95 \\
6 & ETH Zurich & \cellcolor{blue!15}RS & 213 & 8064 & 2.64 \\
7 & Carnegie Mellon University & \cellcolor{blue!15}RS & 211 & 8064 & 2.62 \\
8 & Max Planck Institute for Intelligent Systems & \cellcolor{blue!15}RS & 170 & 8064 & 2.11 \\
9 & Texas A\&M University–Kingsville & \cellcolor{orange!15}RW & 116 & 8064 & 1.44 \\
10 & University of Rostock & \cellcolor{orange!15}RW & 98 & 8064 & 1.22 \\
11 & Henan University & \cellcolor{orange!15}RW & 96 & 8064 & 1.19 \\
12 & Dong A University & \cellcolor{orange!15}RW & 89 & 8064 & 1.10 \\
13 & University of Lagos & \cellcolor{orange!15}RW & 72 & 8064 & 0.89 \\
14 & University of Gondar & \cellcolor{orange!15}RW & 67 & 8064 & 0.83 \\
15 & Midlands State University & \cellcolor{orange!15}RW & 60 & 8064 & 0.74 \\
16 & Savannah State University & \cellcolor{orange!15}RW & 44 & 8064 & 0.55 \\
\bottomrule
\end{tabular}
\caption{Affiliation win rates for \textbf{Meta-Llama-3.1-70B-Instruct}.}
\label{tab:winrate-metallama3170binstruct}
\end{table*}
\begin{table*}[htbp]
\centering
\begin{tabular}{r l l r r r}
\toprule
Rank & Affiliation & Type & Wins & Matches & Win (\%) \\
\midrule
1 & MIT & \cellcolor{blue!15}RS & 478 & 8064 & 5.93 \\
2 & Max Planck Institute for Intelligent Systems & \cellcolor{blue!15}RS & 451 & 8064 & 5.59 \\
3 & ETH Zurich & \cellcolor{blue!15}RS & 440 & 8064 & 5.46 \\
4 & Tsinghua University & \cellcolor{blue!15}RS & 418 & 8064 & 5.18 \\
5 & Carnegie Mellon University & \cellcolor{blue!15}RS & 398 & 8064 & 4.94 \\
6 & University of Cambridge & \cellcolor{blue!15}RS & 391 & 8064 & 4.85 \\
7 & Peking University & \cellcolor{blue!15}RS & 350 & 8064 & 4.34 \\
8 & TU Munich & \cellcolor{blue!15}RS & 340 & 8064 & 4.22 \\
9 & Texas A\&M University–Kingsville & \cellcolor{orange!15}RW & 187 & 8064 & 2.32 \\
10 & University of Rostock & \cellcolor{orange!15}RW & 154 & 8064 & 1.91 \\
11 & Dong A University & \cellcolor{orange!15}RW & 149 & 8064 & 1.85 \\
12 & Midlands State University & \cellcolor{orange!15}RW & 148 & 8064 & 1.84 \\
13 & Henan University & \cellcolor{orange!15}RW & 128 & 8064 & 1.59 \\
14 & Savannah State University & \cellcolor{orange!15}RW & 121 & 8064 & 1.50 \\
15 & University of Lagos & \cellcolor{orange!15}RW & 113 & 8064 & 1.40 \\
16 & University of Gondar & \cellcolor{orange!15}RW & 72 & 8064 & 0.89 \\
\bottomrule
\end{tabular}
\caption{Affiliation win rates for \textbf{Ministral-8B-Instruct-2410}.}
\label{tab:winrate-ministral8binstruct2410}
\end{table*}
\begin{table*}[htbp]
\centering
\begin{tabular}{r l l r r r}
\toprule
Rank & Affiliation & Type & Wins & Matches & Win (\%) \\
\midrule
1 & University of Cambridge & \cellcolor{blue!15}RS & 1316 & 8064 & 16.32 \\
2 & MIT & \cellcolor{blue!15}RS & 1282 & 8064 & 15.90 \\
3 & Carnegie Mellon University & \cellcolor{blue!15}RS & 1230 & 8064 & 15.25 \\
4 & Max Planck Institute for Intelligent Systems & \cellcolor{blue!15}RS & 1229 & 8064 & 15.24 \\
5 & ETH Zurich & \cellcolor{blue!15}RS & 1108 & 8064 & 13.74 \\
6 & Peking University & \cellcolor{blue!15}RS & 1021 & 8064 & 12.66 \\
7 & TU Munich & \cellcolor{blue!15}RS & 1018 & 8064 & 12.62 \\
8 & Tsinghua University & \cellcolor{blue!15}RS & 931 & 8064 & 11.55 \\
9 & University of Rostock & \cellcolor{orange!15}RW & 492 & 8064 & 6.10 \\
10 & Texas A\&M University–Kingsville & \cellcolor{orange!15}RW & 457 & 8064 & 5.67 \\
11 & Henan University & \cellcolor{orange!15}RW & 436 & 8064 & 5.41 \\
12 & Dong A University & \cellcolor{orange!15}RW & 420 & 8064 & 5.21 \\
13 & Savannah State University & \cellcolor{orange!15}RW & 415 & 8064 & 5.15 \\
14 & University of Lagos & \cellcolor{orange!15}RW & 372 & 8064 & 4.61 \\
15 & Midlands State University & \cellcolor{orange!15}RW & 306 & 8064 & 3.79 \\
16 & University of Gondar & \cellcolor{orange!15}RW & 269 & 8064 & 3.34 \\
\bottomrule
\end{tabular}
\caption{Affiliation win rates for \textbf{Mistral-Small-Instruct-2409}.}
\label{tab:winrate-mistralsmallinstruct2409}
\end{table*}
\begin{table*}[htbp]
\centering
\begin{tabular}{r l l r r r}
\toprule
Rank & Affiliation & Type & Wins & Matches & Win (\%) \\
\midrule
1 & Max Planck Institute for Intelligent Systems & \cellcolor{blue!15}RS & 2095 & 8064 & 25.98 \\
2 & MIT & \cellcolor{blue!15}RS & 1989 & 8064 & 24.67 \\
3 & TU Munich & \cellcolor{blue!15}RS & 1814 & 8064 & 22.50 \\
4 & Tsinghua University & \cellcolor{blue!15}RS & 1781 & 8064 & 22.09 \\
5 & University of Cambridge & \cellcolor{blue!15}RS & 1742 & 8064 & 21.60 \\
6 & ETH Zurich & \cellcolor{blue!15}RS & 1728 & 8064 & 21.43 \\
7 & Carnegie Mellon University & \cellcolor{blue!15}RS & 1683 & 8064 & 20.87 \\
8 & Peking University & \cellcolor{blue!15}RS & 1573 & 8064 & 19.51 \\
9 & University of Rostock & \cellcolor{orange!15}RW & 910 & 8064 & 11.28 \\
10 & University of Lagos & \cellcolor{orange!15}RW & 832 & 8064 & 10.32 \\
11 & Texas A\&M University–Kingsville & \cellcolor{orange!15}RW & 819 & 8064 & 10.16 \\
12 & Dong A University & \cellcolor{orange!15}RW & 789 & 8064 & 9.78 \\
13 & Midlands State University & \cellcolor{orange!15}RW & 765 & 8064 & 9.49 \\
14 & University of Gondar & \cellcolor{orange!15}RW & 760 & 8064 & 9.42 \\
15 & Henan University & \cellcolor{orange!15}RW & 726 & 8064 & 9.00 \\
16 & Savannah State University & \cellcolor{orange!15}RW & 672 & 8064 & 8.33 \\
\bottomrule
\end{tabular}
\caption{Affiliation win rates for \textbf{QwQ-32B}.}
\label{tab:winrate-qwq32b}
\end{table*}
\begin{table*}[htbp]
\centering
\begin{tabular}{r l l r r r}
\toprule
Rank & Affiliation & Type & Wins & Matches & Win (\%) \\
\midrule
1 & MIT & \cellcolor{blue!15}RS & 1518 & 8064 & 18.82 \\
2 & Max Planck Institute for Intelligent Systems & \cellcolor{blue!15}RS & 1358 & 8064 & 16.84 \\
3 & TU Munich & \cellcolor{blue!15}RS & 1332 & 8064 & 16.52 \\
4 & ETH Zurich & \cellcolor{blue!15}RS & 1248 & 8064 & 15.48 \\
5 & Carnegie Mellon University & \cellcolor{blue!15}RS & 1204 & 8064 & 14.93 \\
6 & Peking University & \cellcolor{blue!15}RS & 1158 & 8064 & 14.36 \\
7 & University of Cambridge & \cellcolor{blue!15}RS & 1120 & 8064 & 13.89 \\
8 & Tsinghua University & \cellcolor{blue!15}RS & 1041 & 8064 & 12.91 \\
9 & University of Rostock & \cellcolor{orange!15}RW & 769 & 8064 & 9.54 \\
10 & University of Lagos & \cellcolor{orange!15}RW & 573 & 8064 & 7.11 \\
11 & Texas A\&M University–Kingsville & \cellcolor{orange!15}RW & 548 & 8064 & 6.80 \\
12 & Henan University & \cellcolor{orange!15}RW & 507 & 8064 & 6.29 \\
13 & Midlands State University & \cellcolor{orange!15}RW & 492 & 8064 & 6.10 \\
14 & Savannah State University & \cellcolor{orange!15}RW & 467 & 8064 & 5.79 \\
15 & Dong A University & \cellcolor{orange!15}RW & 457 & 8064 & 5.67 \\
16 & University of Gondar & \cellcolor{orange!15}RW & 431 & 8064 & 5.34 \\
\bottomrule
\end{tabular}
\caption{Affiliation win rates for \textbf{GPT-4o-Mini}.}
\label{tab:winrate-gpt4omini}
\end{table*}


\begin{table*}[htbp]
\centering
\resizebox{\textwidth}{!}{
\begin{tabular}{l l l c c c}
\toprule
\textbf{Model} & \textbf{Affiliation} & \textbf{Label} &
\textbf{100 TTP $>$ 0 TTP (\%)} &
\textbf{0 TTP $>$ 100 TTP (\%)} &
\textbf{Tie (\%)} \\
\midrule

\multirow{4}{*}{\makecell[c]{Ministral 8B}}
    & \multirow{2}{*}{RS}
        & Accepted & \textbf{\textcolor{blue}{7.9}} & 3.2 & 88.9 \\
    &   & Rejected & \textbf{\textcolor{blue}{6.3}} & 0.0 & 93.7 \\
\cmidrule(lr){2-6}

    & \multirow{2}{*}{RW}
        & Accepted & \textbf{\textcolor{blue}{11.1}} & 1.6 & 87.3 \\
    &   & Rejected & \textbf{\textcolor{blue}{11.1}} & 3.2 & 85.7 \\
\midrule

\multirow{4}{*}{\makecell[c]{DeepSeek R1\\Distill Llama 8B}}
    & \multirow{2}{*}{RS}
        & Accepted & \textbf{\textcolor{blue}{20.6}} & 6.3 & 73.1 \\
    &   & Rejected & \textbf{\textcolor{blue}{25.4}} & 3.2 & 71.4 \\
\cmidrule(lr){2-6}

    & \multirow{2}{*}{RW}
        & Accepted & \textbf{\textcolor{blue}{28.6}} & 4.8 & 66.6 \\
    &   & Rejected & \textbf{\textcolor{blue}{15.9}} & 4.8 & 79.3 \\
\midrule

\multirow{4}{*}{\makecell[c]{Llama 3.1 8B}}
    & \multirow{2}{*}{RS}
        & Accepted & \textbf{\textcolor{blue}{12.7}} & 0.0 & 87.3 \\
    &   & Rejected & \textbf{\textcolor{blue}{14.3}} & 0.0 & 85.7 \\
\cmidrule(lr){2-6}

    & \multirow{2}{*}{RW}
        & Accepted & \textbf{\textcolor{blue}{9.5}} & 0.0 & 90.5 \\
    &   & Rejected & \textbf{\textcolor{blue}{11.1}} & 3.2 & 85.7 \\
\midrule

\multirow{4}{*}{\makecell[c]{Mistral Small 22B}}
    & \multirow{2}{*}{RS}
        & Accepted & \textbf{\textcolor{blue}{31.7}} & 1.6 & 66.7 \\
    &   & Rejected & \textbf{\textcolor{blue}{33.3}} & 0.0 & 66.7 \\
\cmidrule(lr){2-6}

    & \multirow{2}{*}{RW}
        & Accepted & \textbf{\textcolor{blue}{42.9}} & 3.2 & 53.9 \\
    &   & Rejected & \textbf{\textcolor{blue}{38.1}} & 1.6 & 60.3 \\
\midrule

\multirow{4}{*}{\makecell[c]{DeepSeek R1\\Distill Qwen 32B}}
    & \multirow{2}{*}{RS}
        & Accepted & \textbf{\textcolor{blue}{20.6}} & 6.3 & 73.1 \\
    &   & Rejected & \textbf{\textcolor{blue}{19.0}} & 3.2 & 77.8 \\
\cmidrule(lr){2-6}

    & \multirow{2}{*}{RW}
        & Accepted & \textbf{\textcolor{blue}{28.6}} & 4.8 & 66.6 \\
    &   & Rejected & \textbf{\textcolor{blue}{15.9}} & 6.3 & 77.8 \\
\midrule

\multirow{4}{*}{\makecell[c]{QwQ 32B}}
    & \multirow{2}{*}{RS}
        & Accepted & \textbf{\textcolor{blue}{42.9}} & 4.8 & 52.3 \\
    &   & Rejected & \textbf{\textcolor{blue}{52.4}} & 1.6 & 46.0 \\
\cmidrule(lr){2-6}

    & \multirow{2}{*}{RW}
        & Accepted & \textbf{\textcolor{blue}{38.1}} & 1.6 & 60.3 \\
    &   & Rejected & \textbf{\textcolor{blue}{44.4}} & 3.2 & 52.4 \\
\midrule

\multirow{4}{*}{\makecell[c]{Llama 3.1\\70B Instruct}}
    & \multirow{2}{*}{RS}
        & Accepted & \textbf{\textcolor{blue}{17.5}} & 1.6 & 80.9 \\
    &   & Rejected & \textbf{\textcolor{blue}{19.0}} & 0.0 & 81.0 \\
\cmidrule(lr){2-6}

    & \multirow{2}{*}{RW}
        & Accepted & \textbf{\textcolor{blue}{19.0}} & 0.0 & 81.0 \\
    &   & Rejected & \textbf{\textcolor{blue}{20.6}} & 0.0 & 79.4 \\
\midrule

\multirow{4}{*}{\makecell[c]{Gemini 2.0\\Flash Lite}}
    & \multirow{2}{*}{RS}
        & Accepted & \textbf{\textcolor{blue}{31.7}} & 11.1 & 57.2 \\
    &   & Rejected & \textbf{\textcolor{blue}{33.3}} & 1.6 & 65.1 \\
\cmidrule(lr){2-6}

    & \multirow{2}{*}{RW}
        & Accepted & \textbf{\textcolor{blue}{28.6}} & 7.9 & 63.5 \\
    &   & Rejected & \textbf{\textcolor{blue}{23.8}} & 9.5 & 66.7 \\
\midrule

\multirow{4}{*}{\makecell[c]{GPT-4o Mini}}
    & \multirow{2}{*}{RS}
        & Accepted & \textbf{\textcolor{blue}{42.9}} & 0.0 & 57.1 \\
    &   & Rejected & \textbf{\textcolor{blue}{34.9}} & 0.0 & 65.1 \\
\cmidrule(lr){2-6}

    & \multirow{2}{*}{RW}
        & Accepted & \textbf{\textcolor{blue}{52.4}} & 0.0 & 47.6 \\
    &   & Rejected & \textbf{\textcolor{blue}{34.9}} & 1.6 & 63.5 \\
\midrule

\end{tabular}
}
\label{tab:pubhistory}
\caption{\textbf{Publication history bias.} \% of papers where the LLM assigns a hard higher rating to the author shown with 100 top-tier publications (TTP) compared to 0 TTP. \textbf{\textcolor{blue}{Blue}} indicates the higher value in each pair.}
\end{table*}

\begin{table*}[htbp]
\centering
\resizebox{\textwidth}{!}{
\begin{tabular}{l l l c c c}
\toprule
\textbf{Model} & \textbf{Affiliation} & \textbf{Label} &
\textbf{Senior PI $>$ UG (\%)} & \textbf{UG $>$ Senior PI (\%)} & \textbf{Tie (\%)} \\
\midrule

\multirow{4}{*}{\makecell[c]{Ministral 8B}}
    & \multirow{2}{*}{RS}
        & Accepted & \textbf{\textcolor{blue}{6.3}} & 0.0 & 93.7 \\
    &   & Rejected & \textbf{\textcolor{blue}{14.3}} & 0.0 & 85.7 \\
\cmidrule(lr){2-6}

    & \multirow{2}{*}{RW}
        & Accepted & \textbf{\textcolor{blue}{11.1}} & 0.0 & 88.9 \\
    &   & Rejected & \textbf{\textcolor{blue}{23.8}} & 1.6 & 74.6 \\
\midrule

\multirow{4}{*}{\makecell[c]{DeepSeek R1\\Distill Llama 8B}}
    & \multirow{2}{*}{RS}
        & Accepted & \textbf{\textcolor{blue}{15.9}} & 9.5 & 74.6 \\
    &   & Rejected & \textbf{\textcolor{blue}{22.2}} & 4.8 & 73.0 \\
\cmidrule(lr){2-6}

    & \multirow{2}{*}{RW}
        & Accepted & \textbf{\textcolor{blue}{20.6}} & 6.3 & 73.1 \\
    &   & Rejected & \textbf{\textcolor{blue}{15.9}} & 7.9 & 76.2 \\
\midrule

\multirow{4}{*}{\makecell[c]{Llama 3.1 8B}}
    & \multirow{2}{*}{RS}
        & Accepted & \textbf{\textcolor{blue}{9.5}} & 3.2 & 87.3 \\
    &   & Rejected & \textbf{\textcolor{blue}{7.9}} & 4.8 & 87.3 \\
\cmidrule(lr){2-6}

    & \multirow{2}{*}{RW}
        & Accepted & \textbf{\textcolor{blue}{11.1}} & 1.6 & 87.3 \\
    &   & Rejected & \textbf{\textcolor{blue}{15.9}} & 3.2 & 80.9 \\
\midrule

\multirow{4}{*}{\makecell[c]{Mistral Small 22B}}
    & \multirow{2}{*}{RS}
        & Accepted & \textbf{\textcolor{blue}{25.4}} & 3.2 & 71.4 \\
    &   & Rejected & \textbf{\textcolor{blue}{22.2}} & 1.6 & 76.2 \\
\cmidrule(lr){2-6}

    & \multirow{2}{*}{RW}
        & Accepted & \textbf{\textcolor{blue}{38.1}} & 0.0 & 61.9 \\
    &   & Rejected & \textbf{\textcolor{blue}{44.4}} & 0.0 & 55.6 \\
\midrule

\multirow{4}{*}{\makecell[c]{DeepSeek R1\\Distill Qwen 32B}}
    & \multirow{2}{*}{RS}
        & Accepted & \textbf{\textcolor{blue}{15.9}} & 4.8 & 79.3 \\
    &   & Rejected & \textbf{\textcolor{blue}{15.9}} & 7.9 & 76.2 \\
\cmidrule(lr){2-6}

    & \multirow{2}{*}{RW}
        & Accepted & \textbf{\textcolor{blue}{34.9}} & 7.9 & 57.2 \\
    &   & Rejected & \textbf{\textcolor{blue}{17.5}} & 14.3 & 68.2 \\
\midrule

\multirow{4}{*}{\makecell[c]{QwQ 32B}}
    & \multirow{2}{*}{RS}
        & Accepted & \textbf{\textcolor{blue}{27.0}} & 6.3 & 66.7 \\
    &   & Rejected & \textbf{\textcolor{blue}{31.7}} & 7.9 & 60.4 \\
\cmidrule(lr){2-6}

    & \multirow{2}{*}{RW}
        & Accepted & \textbf{\textcolor{blue}{27.0}} & 7.9 & 65.1 \\
    &   & Rejected & \textbf{\textcolor{blue}{30.2}} & 4.8 & 65.0 \\
\midrule

\multirow{4}{*}{\makecell[c]{Llama 3.1\\70B Instruct}}
    & \multirow{2}{*}{RS}
        & Accepted & 1.6 & 1.6 & 96.8 \\
    &   & Rejected & \textbf{\textcolor{blue}{7.9}} & 0.0 & 92.1 \\
\cmidrule(lr){2-6}

    & \multirow{2}{*}{RW}
        & Accepted & \textbf{\textcolor{blue}{6.3}} & 0.0 & 93.7 \\
    &   & Rejected & \textbf{\textcolor{blue}{9.5}} & 0.0 & 90.5 \\
\midrule

\multirow{4}{*}{\makecell[c]{Gemini 2.0\\Flash Lite}}
    & \multirow{2}{*}{RS}
        & Accepted & \textbf{\textcolor{blue}{41.3}} & 4.8 & 53.9 \\
    &   & Rejected & \textbf{\textcolor{blue}{39.7}} & 3.2 & 57.1 \\
\cmidrule(lr){2-6}

    & \multirow{2}{*}{RW}
        & Accepted & \textbf{\textcolor{blue}{46.0}} & 6.3 & 47.7 \\
    &   & Rejected & \textbf{\textcolor{blue}{47.6}} & 1.6 & 50.8 \\
\midrule

\multirow{4}{*}{\makecell[c]{GPT-4o Mini}}
    & \multirow{2}{*}{RS}
        & Accepted & \textbf{\textcolor{blue}{23.8}} & 6.3 & 69.9 \\
    &   & Rejected & \textbf{\textcolor{blue}{14.3}} & 3.2 & 82.5 \\
\cmidrule(lr){2-6}

    & \multirow{2}{*}{RW}
        & Accepted & \textbf{\textcolor{blue}{31.7}} & 3.2 & 65.1 \\
    &   & Rejected & \textbf{\textcolor{blue}{19.0}} & 3.2 & 77.8 \\
\midrule

\end{tabular}
}
\caption{\textbf{Seniority bias}. \% of papers where the LLM assigns a higher hard rating to a Senior PI profile compared to an Undergraduate Student (UG). \textbf{\textcolor{blue}{Blue}} indicates the higher value in each pair.}
\label{tab:senioritybias}
\end{table*}

\end{document}